\documentclass[aps, prb,10pt,twocolumn,amsmath,amssymb, %showpacs,
                 superscriptaddress,longbibliography]{revtex4-1}
\usepackage[]{graphicx}
\usepackage[]{verbatim}
\usepackage[dvipsnames]{xcolor}
\usepackage{tabularx}

\usepackage{amsfonts}
\usepackage{amsmath}
\usepackage{amssymb}
\usepackage{bm}
\usepackage{graphicx}
\usepackage[breaklinks,colorlinks=true,citecolor=blue]{hyperref}
\usepackage{mathrsfs}
\usepackage[lofdepth,lotdepth,caption=false]{subfig}
\usepackage{varwidth}
\usepackage{wrapfig}
\usepackage{times}
\usepackage{longtable}
\usepackage{multirow}
\usepackage{tikz}
\usepackage{mathtools}

\begin{document}

%\preprint{APS/123-QED}

\title{Anomalous Hall signatures of nonsymmorphic nodal lines in doped chromium chalcospinel CuCr$_2$Se$_4$}

\author{Subhasis Samanta}
\affiliation{Department of Physics, Kangwon National University, Chuncheon 24341, Korea}

\author{Gang Chen}
\email{gangchen@hku.hk}
\affiliation{Department of Physics and Center of Theoretical and Computational Physics,
The University of Hong Kong, Pokfulam Road, Hong Kong, China}
\affiliation{State Key Laboratory of Surface Physics and Department of Physics, Fudan University, Shanghai 200433, China}

\author{Heung-Sik Kim}
\email{heungsikim@kangwon.ac.kr}
\affiliation{Department of Physics, Kangwon National University, Chuncheon 24341, Korea}
\affiliation{Institute for Accelerator Science, Kangwon National University, Chuncheon 24341, Korea}

\begin{abstract}
An emerging phase of matter among the class of topological materials is nodal line semimetal, possessing symmetry-protected one-dimensional gapless lines at the (or close to) the Fermi level in $k$-space. When the $k$-dispersion of the nodal line is weak, van Hove singularities generated by the almost flat nodal lines may be prone to instabilities introduced by additional perturbations such as spin-orbit coupling or magnetism. Here, we study Cr-based ferromagnetic chalcospinel compound CuCr$_2$Se$_4$ (CCS) via first-principles electronic structure methods and reveal the true origin of its dissipationless anomalous Hall conductivity, which was not well understood previously. We find that CCS hosts nodal lines protected by nonsymmorphic symmetries, located in the vicinity of Fermi level, and that such nodal lines are the origin of the previously observed distinct behavior of the anomalous Hall signature in the presence of electron doping. The splitting of nodal line via spin-orbit coupling produces a large Berry curvature, which leads to a significant response in anomalous Hall conductivity. Upon electron doping via chemical substitution or gating, or rotation of magnetization via external magnetic field, noticeable change of anomalous Hall behavior occurs, which makes CCS a promising compound for low energy spintronics applications.
\end{abstract}

\maketitle

\section{Introduction}

Dissipationless charge or spin transport has been a key concept in condensed matter physics because of realizations of low-power electronic and spintronic devices \cite{Gilbert2021,Fan2016,Chang2015} and long-time preservation of quantum information that it promises \cite{Murakami2003,Culcer2020,He2018,Tokura2019}. Dissipationless transport phenomena often originates from topology of electronic structures, and it has been a hallmark of various topologically nontrivial states of matter such as quantum anomalous and spin Hall phases \cite{Nagaosa2010,Kane2010}, or topological semimetals such as magnetic Weyl and nodal semimetals \cite{Vishwanath2011,Ran2011}.

Magnetic topological material, such as topological insulators doped with magnetic ions \cite{Yu2010,Chang2013,Moodera2015} or magnetic nodal semimetals \cite{Zhong2011,Bernevig2020,Higo2015,Kim2018}, are promising candidate to realize dissipationless Hall transport in room-temperature conditions. In terms of practical applications, on the other hand, vanishing density of states in Weyl or Dirac semimetals at the Fermi level is not favorable for good transport properties. In this regard nodal line semimetals with weak momentum space dispersion, which can host van Hove singularities in the vicinity of Fermi level, thanks to the one-dimensional line of zero-energy modes with nonvanishing measure in the $k$-space, seems promising in realizing dissipationless Hall transport or engineering various instabilities for practical device applications\cite{Yang2017,Sun2017,Yang2018}.

It was previously reported that a chalcospinel compound, CuCr$_2$Se$_{4}$ (CCS) doped with Br, shows dissipationless anomalous Hall transport that is unaffected by the doping-induced disorder \cite{Ong2004,Cava2004}. Soon after the experimental report it was suggested to be an electronic Berry phase effect \cite{Niu2006,Zhong2007}. However, the origin of the sharp doping-induced sign change in the anomalous Hall conductivity, as observed both in experiment \cite{Ong2004} and theory \cite{Zhong2007}, has not been understood well. In light of recent advancements in topological band theory, we revisit the anomalous Hall response of the electron-doped CCS and reveal its origin. We found that the Br-doped CCS in its ferromagnetic state ($T_C \simeq 430$ K) is a half-metallic nodal line semimetal, where the nodal degeneracy at the zone boundary is protected by nonsymmorphic symmetries in the absence of spin-orbit coupling (SOC). Inclusion of Cr and Se SOC slightly splits the nodal degeneracy, generating a substantial sign-changing anomalous Hall response under electron doping because of the nodal-line-induced finite density of states. It is further shown that, thanks to its weak magnetocrystalline anisotropy \cite{Masumoto1978}, Br-doped CuCr$_2$Se$_{4}$ may show significant field-dependent response of anomalous Hall signatures under external magnetic fields. Therefore it is expected that electron-doped CuCr$_2$Se$_{4}$ would become a promising platform for dissipationless charge and spin transport applications.

\section{Computational Methods}

For this study we carried out electronic structure calculations within local density approximation (LDA) using Vienna {\it Ab-initio} Simulation Package ({\sc VASP}) \cite{Kresse1996}. Detailed investigation of spin-orbit coupling and computation of anomalous Hall conductivity vector was performed using {\it{ab-initio}}-based tight-binding model constructed via Wannier function \cite{Souza2001}  method as implemented in {\sc wannier90} package \cite{Pizzi2020}. Please refer to the Appendix ~\ref{Appendix:Comp} for more details.

\section{Band structure of CCS} 

\begin{center}
\begin{figure}
\includegraphics[angle=-0,origin=c,scale=0.325]{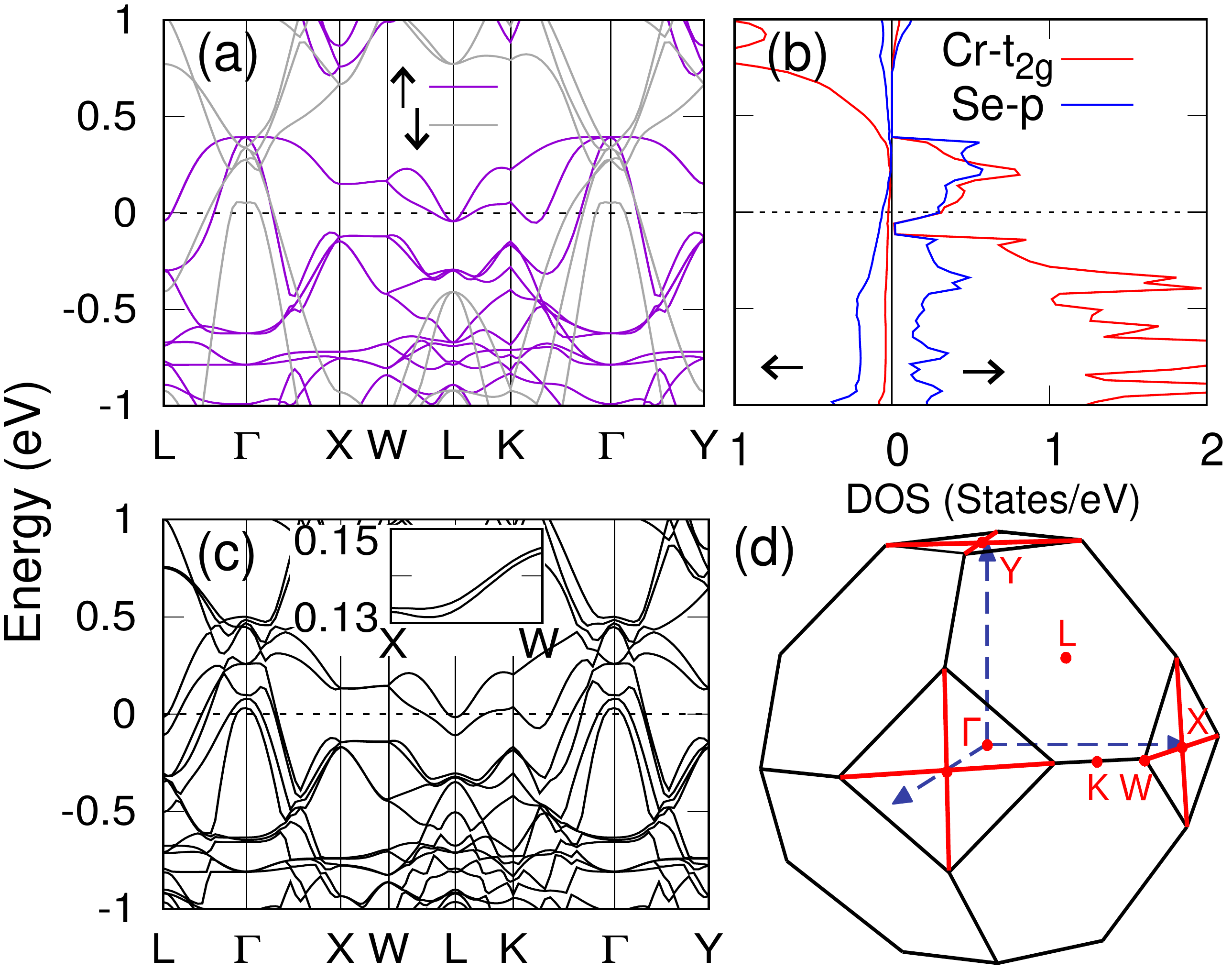}
\caption{Spin-polarized band structure and partial density of states (PDOS), obtained using (a)-(b) LDA and (c) LDA+SOC with magnetization direction parallel to [111]. PDOS clearly shows Van Hove singularities around Fermi level, where nodal lines are seen in the band structure. (d) First Brillouin zone of FCC lattice shows high symmetry points and paths of the nodal lines along $X$-$W$ with thick red lines.}
\label{fig:Fig1}
\end{figure}
\end{center}

CCS is a metallic ferrimagnet ($T_{\rm C} = 430$ K) with a cubic symmetry (space group No. 227), where the ferromagnetic ordering of Cr high-spin moments mostly contributes to the net magnetization \cite{Kanomata2001}. Band structures of CCS with and without spin-orbit coupling are shown in Fig. \ref{fig:Fig1}, where Cr $t_{\rm 2g}$- and Se $p$-like hole characters can be clearly seen in the metallic bands close to the Fermi level. Cu states are most located about 2 eV below the Fermi level and weakly contribute to the metallic states. Note that charge configurations of Cu and Cr are not exactly Cu$^{+1}$ and Cr$^{3+}$, respectively, because of negative charge transfer from Se$^{2+}$ ions to Cr and Cu sites yielding the metallic behavior.

A salient feature of the CCS band structure is the presence of nearly flat two-fold degenerate bands in the majority spin channel (Fig.~\ref{fig:Fig1}(a)), located close to the Fermi level (within the energy window $\vert E - E_{\rm{F}} \vert \leq 0.15$ eV) on the $X$-$W$ lines in the absence of spin-orbit coupling. These flat nodal lines induce van Hove singularities around $\pm$ 0.15 eV, as shown in the density of states plot (right panel in Fig.~\ref{fig:Fig1}(a)), while dispersive minority spin channel almost does not contribute to the states close to the Fermi level (see Fig.~\ref{fig:Fig1}(b)). These line degeneracies, namely the so-called Weyl nodal line features \cite{Yugui2019}, are protected by two non-symmorphic and perpendicular $d$-glide planes that intersect on every $X$-$W$ lines. Anticommutation relations between the two intersecting glide operations on the X-W lines enforce twofold degeneracy, which can be lifted with the loss of the glide symmetries via lattice distortions or spin-orbit coupling plus symmetry-breaking magnetism.

Indeed, Fig.~\ref{fig:Fig1}(c) shows the band structure of CCS with including SOC and net magnetization parallel to the [111] direction. A small but finite splitting of the nodal lines on the $X$-$W$ line is visible. The splitting persists when the magnetization direction is along [001] axis because the loss of glide plane. Orbital-projected density of states close to the nodal lines shows almost equal mixture of Cr $t_{\rm 2g}$ and Se $p$-states (see right panels of Fig.~\ref{fig:Fig1}(b)), so the splitting of the nodal lines can be attributed to the SOC of Cr $d$- and Se $p$-orbitals.

Recently it has been shown that splitting of nodal lines via SOC results in large intrinsic anomalous Hall or spin Hall signature and topological phases \cite{Kane2015,Sun2017,Kim2018,Manna2018,Noky2019,Li2020,Minami2020}. Thanks to the flat dispersion of the nodal line bands and the resulting finite density of states close to the Fermi level, SOC-induced anomalous Hall signatures in CCS should be significant. Furthermore, small SOC-induced splitting of the nodal lines in CCS may result in a sharp change of anomalous Hall character under charge doping as reported previously \cite{Ong2004, Cava2004}.

\section{Wannierized effective model and SOC}

In order to investigate the role of SOC in anomalous Hall responses, we constructed a tight-binding (TB) Hamiltonian via Wannier orbital method \cite{Marzari1997,Souza2001,Weng2009}. Our Wannierized TB model incorporates 12 Cr-$t_{2g}$-like and 24 Se-$p$-like orbitals and faithfully represents nodal lines and other band features of majority spin channel close to the Fermi level (see Appendix A.2 and Fig.~\ref{fig:SM_Fig1} for more detail). Note that the contribution of the minority spin channel to the Hall conductivity should be marginal at most because of its negligibly small density of states (see Fig.~\ref{fig:Fig1}(b)). 

In the majority spin channel, where spin degree of freedom is frozen because of magnetic ordering, SOC behaves as an effective Zeeman field in the orbital sector; $\hat{H}_{\rm SO} \rightarrow \hat{\mathcal{P}}^\dag_{\bf n} \hat{H}_{\rm SO} \hat{\mathcal{P}}_{\bf n} \simeq \frac{\hbar}{2} \sum_\alpha \lambda_\alpha \hat{L}^\alpha_{\bf n}$, where $\alpha$ is index for Cr and Se sites, $\hat{H}_{\rm SO} \equiv \sum_\alpha \lambda_{\alpha}\hat{\bf L}^\alpha \cdot \hat{\bf S}^\alpha$ is original atomic SOC, ${\bf n}$ is the magnetization direction, and $\hat{\mathcal{P}}_{\bf n} \equiv \sum_\alpha \vert \uparrow_{\bf n} \rangle_\alpha \langle \uparrow_{\bf n} \vert_\alpha$ is a projection operator onto the majority spin channel along ${\bf n}$. Hence, SOC on top of nonzero magnetization along $\bf n$ splits degeneracy between nonzero $L_{\rm n}$ states, which consist of nodal gapless lines at zone boundaries. 

\section{Anomalous Hall conductivity}

To investigate anomalous Hall responses with respect to net magnetization direction in CCS, we employed Fukui-Hatsugai-Suzuki method to compute Berry curvature vector $\boldsymbol{\Omega}({\bf k}) \equiv \left( \Omega_{yz}, \Omega_{xz}, \Omega_{xy} \right) ({\bf k})$ and anomalous Hall conductivity (AHC) in a $96 \times 96 \times 96$ discretized $k$-space \cite{Suzuki2005}. Two magnetization directions, [001] and [111] with respect to the cubic axes, were considered. Both Cr ($\lambda_{\rm Cr}$=0.02 eV) and Se ($\lambda_{\rm Se}$=0.05 eV) SOC were included, where values of $\lambda_{\rm Cr}$ and $\lambda_{\rm Se}$ were chosen to best fit the energy splitting of two bands close to the Fermi level at L point. Note that the Berry curvature vector $\boldsymbol{\Omega}({\bf k})$ becomes identically zero at all $k$-point in the absence of SOC because of coexistence of the complex conjugation ({\it i.e.} product of time-reversal and global spin rotation operations) and inversion symmetries.

\begin{figure*}
\centering
\includegraphics[width=0.8\textwidth]{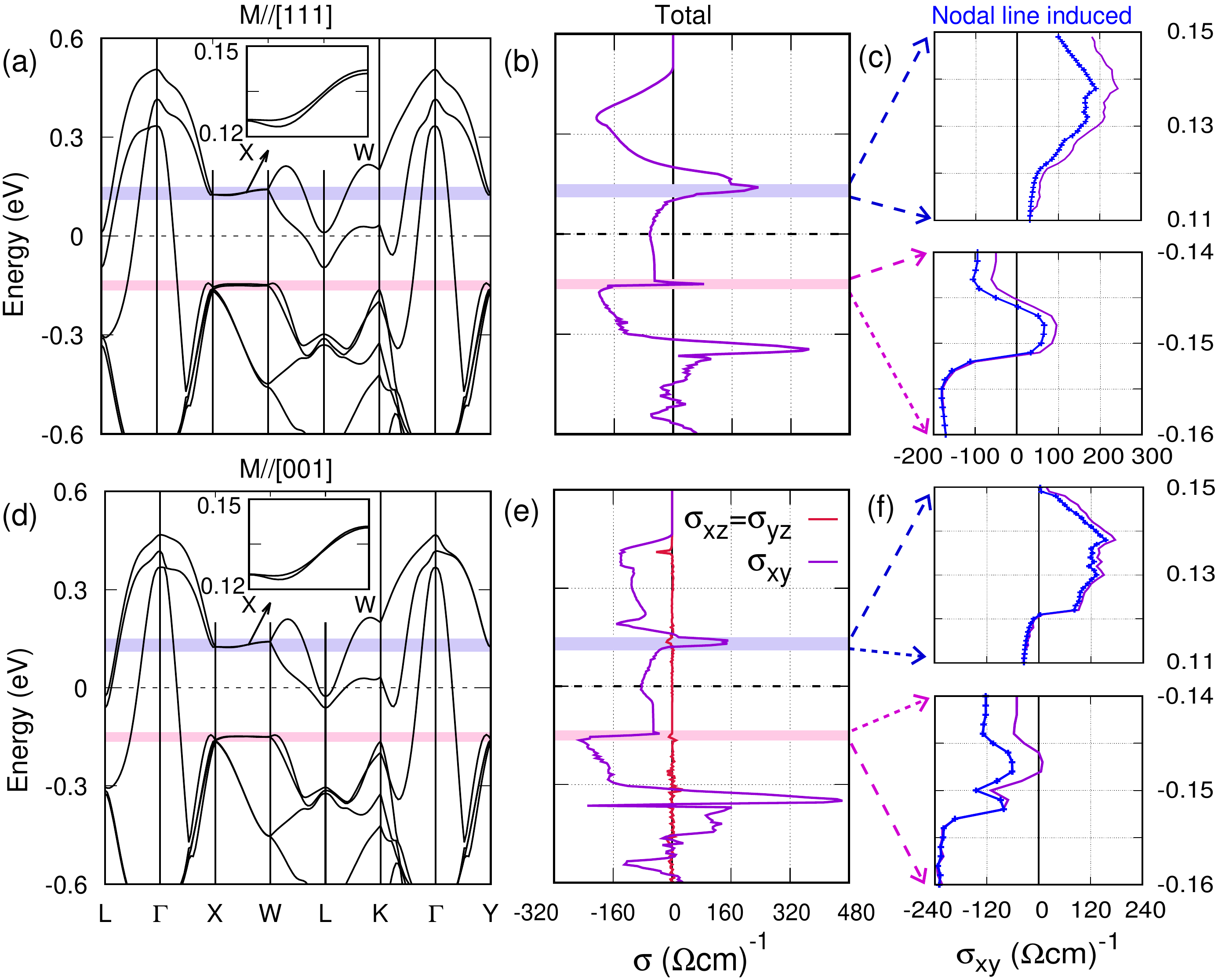}
\caption{(a) Band structure in the majority spin channel, obtained from TB Wannier model with magnetization pointed along [111]. Inset shows the splitting of upper nodal line in the enlarged energy window. (b) Three components of total anomalous Hall conductivities are depicted in violet ($\sigma_{xy}$) and red ($\sigma_{yz,xz}$) colors. A steep increase in Hall conductivity is observed close to the nodal lines. For ${\bf M}//$[111], all three components are identical. (c) AHC induced by upper (top) and lower (bottom) nodal lines (blue dotted line). (d)-(f) Band structure, total AHC, and nodal lines induced AHC, respectively with ${\bf M}//$[001]. The noisy features of $\sigma_{yz,xz}$ in bottom panel, (e) come from numerical noises that vanish in the limit of infinitely dense $k$-grid.}
\label{fig:Fig2}
\end{figure*}

Figure \ref{fig:Fig2} shows the effect of spin-orbit coupling on the Wannier TB band structure with different magnetization directions and the resulting anomalous Hall responses. Here, the inclusion of SOC unlocks two conditions necessary for finite anomalous Hall responses; first, as mentioned above, SOC splits nodal line degeneracies by coupling magnetization and band structure so that glide symmetries are lost except ones perpendicular to the magnetization direction. Second, because the global spin rotation symmetry is lost due to SOC, the complex conjugation symmetry which enforces $\boldsymbol{\Omega}(-{\bf k}) = -\boldsymbol{\Omega}({\bf k})$, is gone. Hence, a finite anomalous Hall response that does not vanish under the $k$-summation emerges from the splitting of the nodal lines via SOC.

The band structures with SOC reveal several important features of the nodal lines. The doubly degenerate nodal lines, which were protected by the nonsymmorphic symmetries in the absence of SOC, are now split via spin-orbit coupling. The insets of Figs. \ref{fig:Fig2} (a) and (d) show SOC induced splitting of upper nodal line. The SOC induced splitting energy is marginal, of the order of few meV. The splitting is dependent on the magnetization direction. In the case of [111], nodal line splitting is more pronounced than the [001] case. The nodal lines also, split at $X$ point, but its magnitude is negligibly small. The direction dependent splitting is more visible for bands close to $E_{\rm F}$ at high symmetry point $L$.

%In the presence of SOC, bands along the high symmetry path $X$-$W$ in Fig. \ref{fig:Fig2} reveals several important features of the nodal line. Here, magnetization direction and spin-orbit coupling constants control the magnitude of the splitting. The SOC induced splitting energy along the nodal lines is marginal, of the order of few meV. In the case of [111] direction, splitting is more pronounced, and also, the role is $\lambda_{\rm Cr}$ is quite clear. The inclusion of $\lambda_{\rm Cr}$ effectively lowers the nodal line splitting. Since nodal line bands consist of hybridized Cr-$t_{2g}$ - Se-$p$ orbitals, net splitting reduces at $W$ point due to the opposite parity of Cr and Se states. The nodal line too splits at $X$ point, but its magnitude is negligibly small. 

\begin{figure}
\begin{center}
\includegraphics[width=0.45\textwidth]{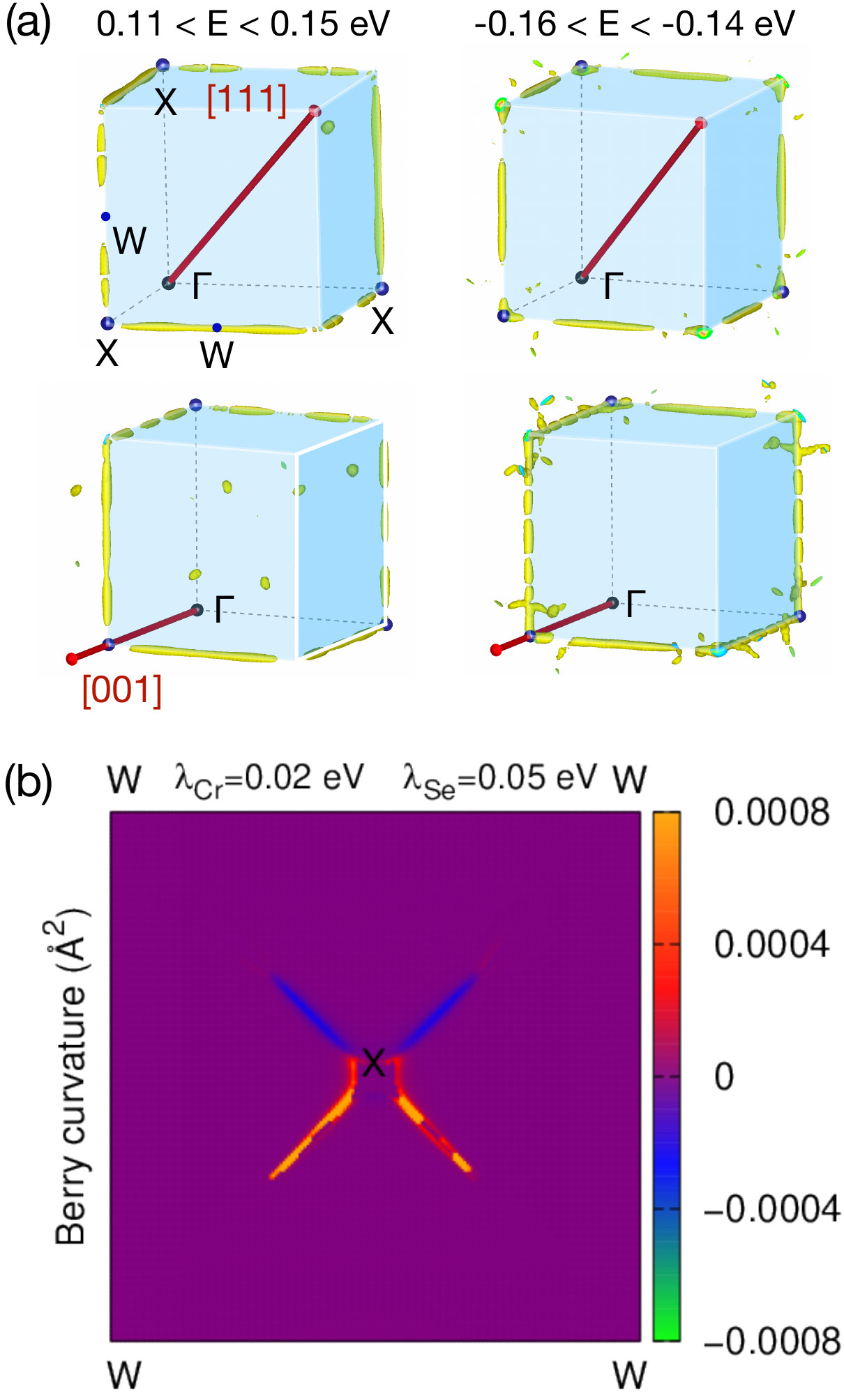}
\caption{
(a) Isosurfaces of magnitudes of energy-integrated Berry curvature vectors, $\left \vert \int^{\mu_2}_{\mu_1} d\mu~ \boldsymbol{\Omega}({\bf k}, \mu) \right \vert$, where the energy range of the integration $\{\mu_1, \mu_2\}$ is chosen to fully include nodal-line-induced Berry curvatures. Left and right panels present results with $\{\mu_1, \mu_2\} = \{0.11, 0.15\}$ and $\{-0.16, -0.14\}$, each of them depicted as violet and pink shades in Fig.~\ref{fig:Fig2}, respectively. Red arrows represent the direction of the magnetization. 
(b) $\Omega_{xy}({\bf k}, \mu)$ on the side Brillouin zone surface containing the $X$-$W$ lines (depicted as blue-shaded planes in (a)), where $\mu$ = 0.125 eV and ${\bf M}//$[111].
}
\label{fig:Fig3}
\end{center}
\end{figure}

Anomalous Hall conductivities $\boldsymbol{\sigma}(\mu)$ as a function of chemical potential $\mu$ is shown in Figs. \ref{fig:Fig2} (b) and (e) alongside band plots, where changes in $\boldsymbol{\sigma}(\mu)$ introduced by the presence of SOC and tilting of magnetization direction from [111] to [001] are shown. Note that anomalous Hall vectors behave in the same way as magnetic moments under symmetry operations, and four-fold and three-fold rotation symmetries remain unbroken in the cases of ${\bf M}//$[001] and [111], respectively. Therefore, components of anomalous Hall vectors that are perpendicular to the magnetization cancel out, and only components parallel to the magnetization directions ($\sigma_{xy}$ when ${\bf M}//$[001], and $\sigma_{xy}=\sigma_{xz}=\sigma_{yz}$ when ${\bf M}//$[111]) survive, as shown in Fig.~\ref{fig:Fig2}. In both cases of ${\bf M}//$[001] and [111], a steep change in Hall conductivity is observed close to the nodal line. Comparing the total AHC for upper nodal line between [111] and [001], we find that former is roughly 80 ($\Omega$cm)$^{-1}$ higher in magnitude than latter one.

From the Fig.~\ref{fig:Fig2}, it is evident that the large enhancement in total AHC occurs close to the nodal lines in the electron doped region. To see the role of nodal lines and also, to quantify its contribution to total AHC, we have extracted the nodal-line-induced contribution to the AHC. Figures~\ref{fig:Fig2}(c) and (f) show upper and lower nodal lines induced AHC (blue dotted lines) and total AHC (solid violet lines) for the $\sigma_{xy}$ component with two magnetization directions. Note that we obtained the nodal-line-induced AHC via sampling $k$-points inside thin slabs enclosing the side surfaces of Brillouin zone that contain the X-W points (slab thickness being 10\% of the reciprocal vectors). The result suggests that most of the contribution to total AHC comes from the nodal lines. This confirms that the previously observed large anomalous Hall current in CCS \cite{Ong2004} in the presence of Br doping actually stems from the splitting of nodal line via spin-orbit coupling.

Furthermore, we plot nodal line-induced Berry curvature in the $k$-space. Figures~\ref{fig:Fig3}(a) show isosurfaces of magnitudes of energy-integrated magnitudes of Berry curvature vectors, namely, $\left \vert \int^{\mu_2}_{\mu_1} d\mu ~ \boldsymbol{\Omega}({\bf k}, \mu) \right \vert$, where $\{\mu_1, \mu_2\}$ is set to be $\{0.11, 0.15\}$ and $\{-0.16, -0.14\}$ eV to enclose the nodal lines above and below the Fermi level, respectively. We also considered both the M // [111] and [001] cases. From Fig.~\ref{fig:Fig3}(a) it can be seen that distribution of the Berry curvature is concentrated on the line connecting the X and W points, so that the sign-changing AHC originates from the splitting of the nodal lines via SOC.

\begin{figure}
\begin{center}
\includegraphics[angle=0,origin=c,scale=0.32]{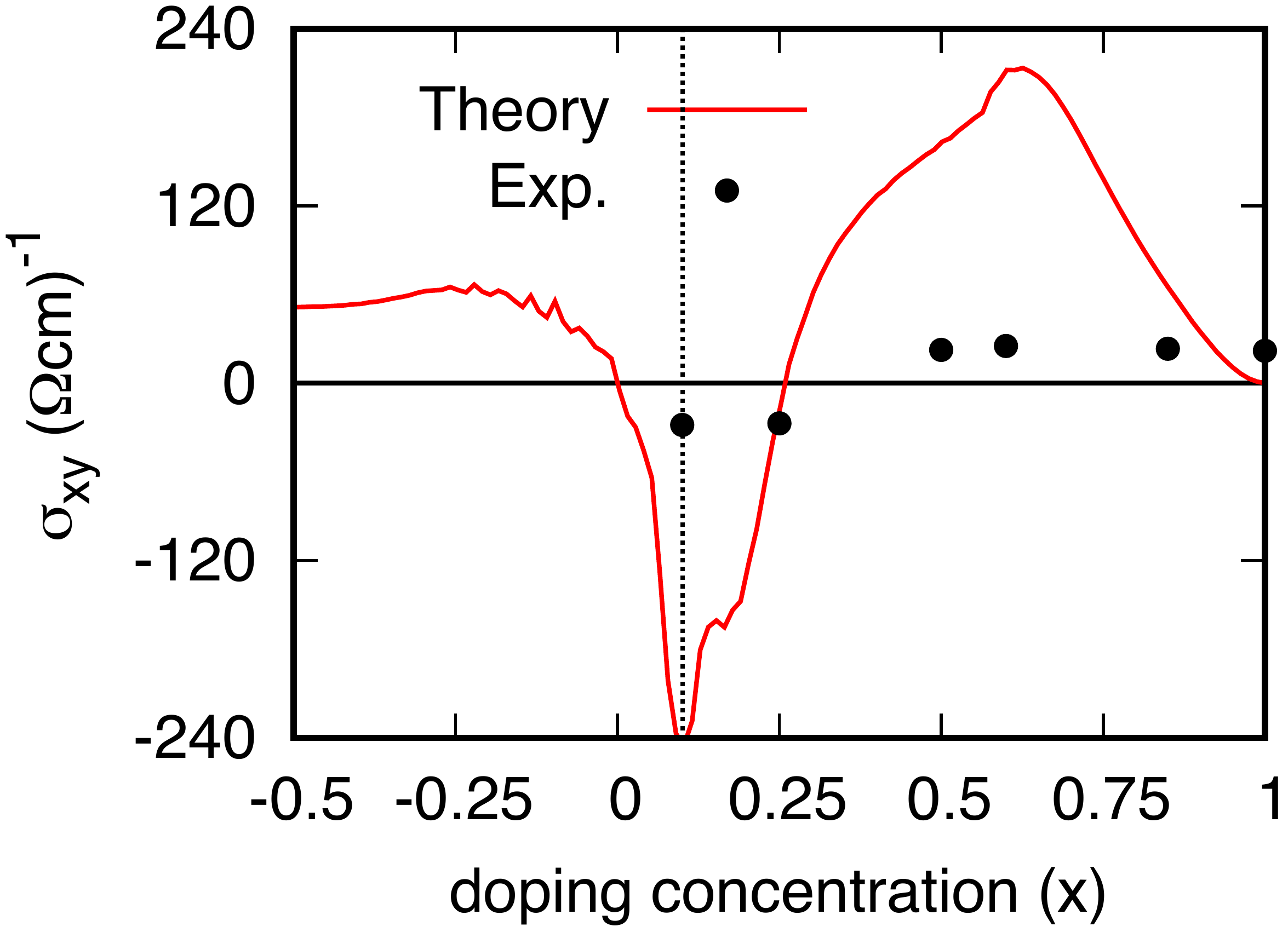}
\caption{Hall conductivity in the presence of nonzero $\lambda_{\rm Cr}$ and $\lambda_{\rm Se}$ when ${\rm M}$//[111], plotted as a function of doping. For comparison, experimental Hall conductivity (measured at 5 K), adopted from the reference \cite{Ong2004} for a few doping concentrations ($x$ = 0.1,0.25,0.5,0.6,0.85,1), are shown with filled circles. Vertical dotted line located around $x=0.1$ marks the position of center of the X-W nodal line dispersion in our calculation (see Fig.~\ref{fig:Fig2}(b).}
\label{fig:Fig4}
\end{center}
\end{figure}

In addition, in Fig.~\ref{fig:Fig3}(b) we plotted the Berry curvature $\Omega_{xy}({\bf k})$ on one square face of the first Brillouin zone containing $X$ and $W$ points when the magnetization is along [111] direction ({\it i.e.} $\Omega_{xy}({\bf k}) = \Omega_{xz}({\bf k}) = \Omega_{yz}({\bf k})$) with $\mu$ = 0.125 eV. Fig. \ref{fig:Fig3}(b) confirms that the Berry curvature is concentrated on the $X$-$W$ nodal lines, as shown in Fig.~\ref{fig:Fig3}(a) and consistent with a previous theoretical finding \cite{Zhong2007}. It is noted that, with the presence of SOC, the loss of time-reversal symmetry causes an imbalance of Berry Curvature at $\textbf{k}$ and -$\textbf{k}$ points with respect to the X point and the resulting nonzero net anomalous Hall conductivity.

Figure \ref{fig:Fig4} shows computed $\sigma_{xy}(x)$ as a function of electron doping per formula unit $x$ when ${\bf M}$//[111]. Previous experimental Hall conductivity \cite{Ong2004} for few selective doping concentrations is also displayed in the figure for comparison. The Hall response is maximum when doping concentration is close to 0.1, where the doping-induced chemical potential crosses the center of the X-W nodal line dispersion. The sign change behavior is captured around $x$=0.25. Origin of experimental $\sigma_{xy}$ close to $x$=1 should be more complicated due to effects from minority-spin bands or impurity scattering, whose contributions are absent in our Wannierized model.

\section{Role of $\lambda_{Cr}$ on band structure and anomalous Hall conductivity}

\begin{figure}
\begin{center}
\includegraphics[angle=-0,origin=c,scale=0.35]{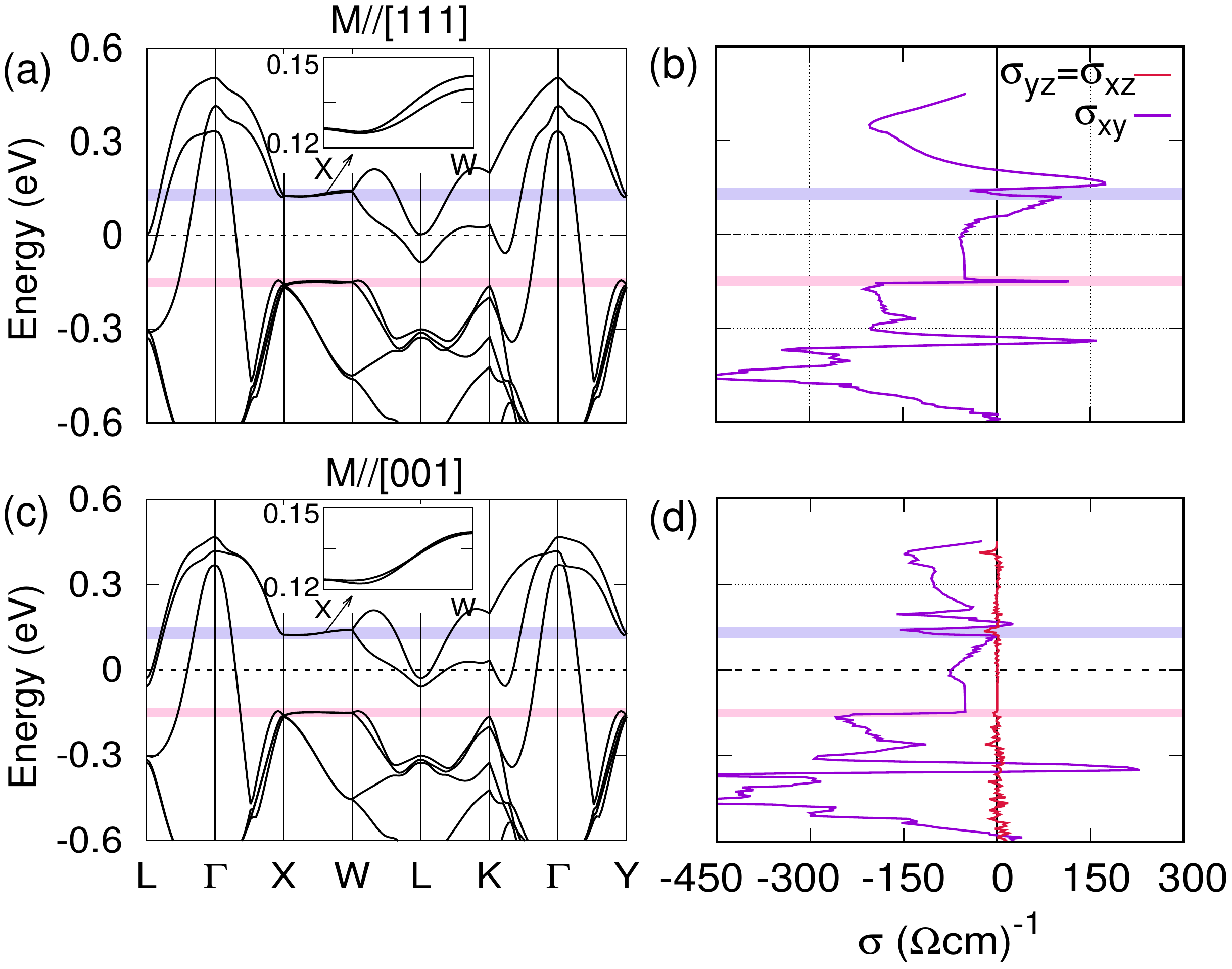}
\caption{(a)-(b) Band structure and total anomalous Hall conductivity, calculated for the majority spin channel with ${\bf M}//$[111] in the presence of $\lambda_{Se}$. (c)-(d) Same band structure and total AHC are now shown for ${\bf M}//$[001]. The shaded regions highlight the nodal lines.}
\label{fig:Fig5}
\end{center}
\end{figure}

In Fig.~\ref{fig:Fig2}, electronic band structure and anomalous Hall conductivity are shown in presence of both Cr and Se SOC. In order to see the role of $\lambda_{\rm Cr}$, we computed band structures and AHC with only including $\lambda_{\rm Se}$ (no $\lambda_{\rm Cr}$), as presented in Fig.~\ref{fig:Fig5}. A close-up view of the upper nodal line splittings without $\lambda_{\rm Cr}$ (insets in Fig.~\ref{fig:Fig5}(a,c)), in comparison to the result with both $\lambda_{\rm Cr}$ and $\lambda_{\rm Se}$ included (insets in Fig.~\ref{fig:Fig2}(a,d)), reveals that for ${\bf M}//$[111], inclusion of $\lambda_{\rm Cr}$ narrows the gap, while there is almost no change of splitting for ${\bf M}//$[001]. However, the effect of $\lambda_{Cr}$ is clearly reflected in the anomalous Hall response as shown in Fig.~\ref{fig:Fig5}(b,d). Since nodal line bands consist of hybridized Cr-$t_{2g}$ - Se-$p$ states, without $\lambda_{\rm Cr}$ AHC which had spiky features at the upper nodal lines, is sharply peaked up with inclusion of $\lambda_{\rm Cr}$ (Fig. \ref{fig:Fig2}). Together this result suggests that both magnetization direction and SOC on both Cr and Se are equally important in order to explain the sign-changing anomalous Hall response close to the nodal lines in the electron-doped regime of this compound.

\section{Potential on-site correlation effects}

In the splitting of nodal lines and resulting anomalous Hall effect in CuCr$_2$Se$_4$, the role of Cr atomic Hund's coupling that gives rise to $S=3/2$ Cr local moments, is crucial. After the Cr-moment-based ferromagnetism occurs, electronic structure calculation results within density functional theory provide a good agreement with experimental observations. However, it always remains to be a valid question whether other parts of Coulomb correlations, such as on-site repulsion, might play a significant role even in metallic systems. To better understand the effect of on-site Coulomb repulsion at Cr sites, we have performed DFT+$U$ \cite{Dudarev1998} and DFT+DMFT (dynamical mean-field theory) \cite{Haule2010,Haule2018} calculations. 

\begin{figure}
\begin{center}
\includegraphics[angle=-0,origin=c,scale=0.6]{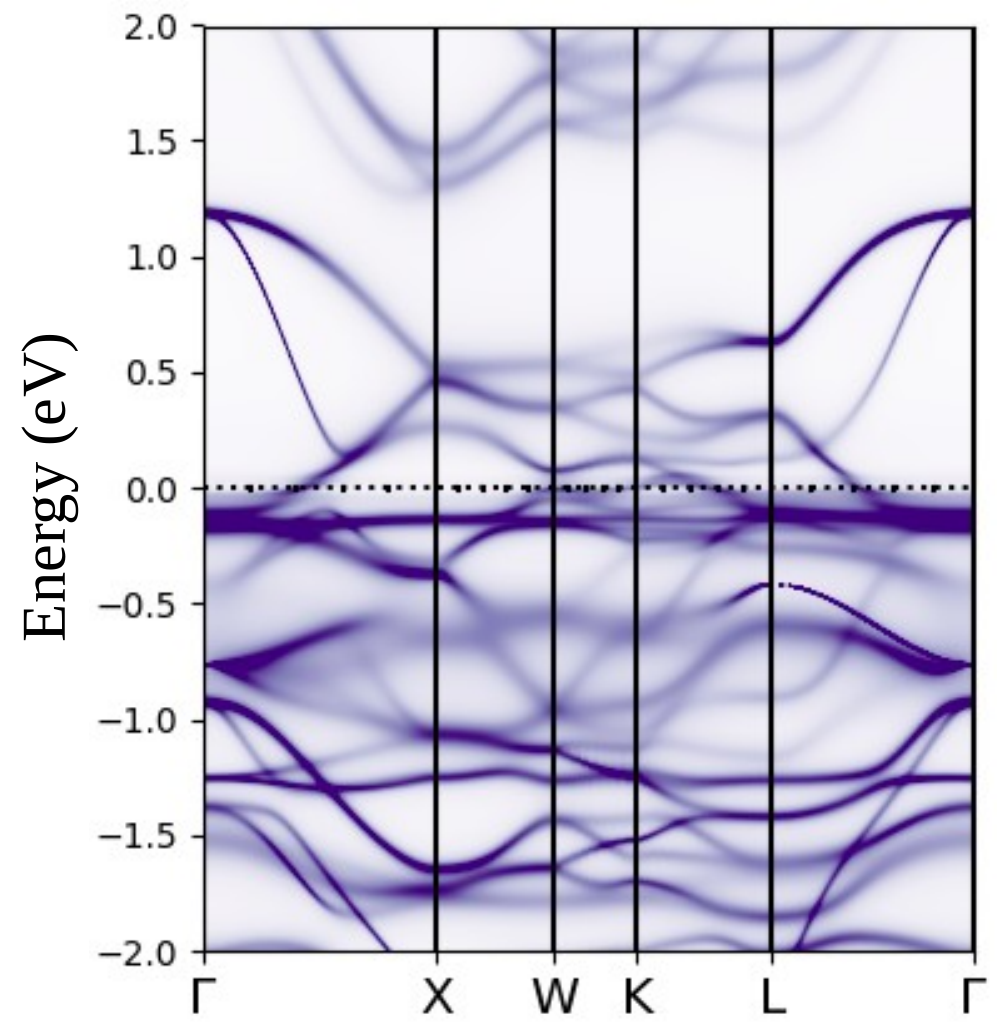}
\caption{A majority-spin spectral function $A(k, \omega)$ obtained from our DMFT calculation.}
\label{fig:SM_Fig4}
\end{center}
\end{figure}

Figure \ref{fig:SM_Fig4} shows a momentum-dependent spectral function $A(k, \omega)$ of majority spin channel of CCS from our DMFT calculation. We notice that majority spin bands are pushed above the Fermi level so that nodal lines lie above the Fermi level within the energy window of 0.4-0.5 eV. $A(k, \omega)$ shows metallic bands ({\it i.e.} clear band-like features without much broadening of spectra) even with the presence of on-site correlations. Top nodal line, located about 0.6 eV above the Fermi level, remains almost dispersionless, while other two nodal lines located about 0 $\sim$ 0.5 eV above the Fermi level are slightly dispersive along the $X$-$W$ line. This shows the robustness of nodal line features and electron-doped anomalous Hall signatures under dynamical electron correlation effects, suggesting that our finding of nodal-line-induced anomalous Hall signatures in doped CuCr$_2$Se$_4$ is relevant to previous experimental findings \cite{Ong2004,Cava2004}.

Interestingly, it is found that DFT+$U$ tends to drive this system away from half-metallic regime, which is inconsistent with experimental observations of strong half-metallicity in this compound \cite{Kanomata2001,Sakurai2007} (see Fig.~\ref{fig:SM_Fig3} and Appendix~\ref{Appendix:HubbardU} for further details), while the the nearly-flat nodal line features remain stable under the effect of dynamic on-site correlations within DFT+DMFT. This origin of this contrasting behavior between DFT+$U$ and DFT+DMFT is unclear at this moment.

\section{Discussion and summary}

The change in magnitude of anomalous Hall conductivity between [001] and [111] near nodal lines reveals that the direction of magnetization also, plays a key role in tuning the Hall response. In addition, according to our density functional theory calculations, energy difference between [001] and [111] FM magnetization directions is found to be very small, 0.098 meV per formula unit (f.u.). This is mostly due to the relatively weak spin-orbit coupling constants at Cr atoms and the basal cubic symmetry. Earlier experimental studies show first and second order anisotropy constants to be of the same order ($K_1$=0.084 and $K_2$=0.011 meV/f.u.), consistent with our finding \cite{Masumoto1978}. Hence, a not-too-strong external magnetic field of less than a Tesla may easily switch the magnetization direction and induce huge anomalous Hall response in this system. 

In conclusion, we explain the origin of large anomalous Hall behavior in the doped CuCr$_2$Se$_{4}$. Our analysis reveals that doubly degenerate nodal line along $X$-$W$ in the vicinity of Fermi-level is responsible for a large Hall signature. In the presence of spin-orbit coupling, splitting of nodal lines produces a non-zero Berry curvature along $X$-$W$, resulting in a large change in anomalous Hall response. In addition, weak magnetocrystalline anisotropy in CuCr$_2$Se$_{4}$ is favorable because it facilitates the switching of the magnetization direction and the resulting anomalous Hall conductivity with the application of an external magnetic field.
%suggests that the switching of magnetization direction is indeed favorable with the application of a rather small magnetic field. 
With these, doped CuCr$_2$Se$_4$ becomes a promising candidate for studying magnetic topological semimetals and realizing low-power dissipationless spintronics applications.

\begin{center}
{\bf ACKNOWLEDGMENTS}
\end{center}

We thank Bohm-Jung Yang for fruitful discussions. S. Samanta was funded by the Korea Research Fellow (KRF) Program of the National Research Foundation of Korea (Grant No. 2019H1D3A1A01102984), and also thanks the National Supercomputing Center of Korea for providing computational resources including technical assistance (Grant No. KSC-2020-CRE-0156). HSK acknowledges the support of the National Research Foundation of Korea (Basic Science Research Program, Grant No. 2020R1C1C1005900). GC was supported by the Ministry of Science and Technology of China
with Grants No. 2018YFE0103200, 2016YFA0300500, 2016YFA0301001,
and by the Shanghai Municipal Science and Technology Major Project with
Grant No. 2019SHZDZX01, and by the Research Grants Council of Hong Kong
with General Research Fund with Grants No. 17303819.

%\begin{center}
%{\bf Author contributions}
%\end{center}

%H.-S. Kim and G. Chen conceived this project. S. Samanta and H.-S. Kim performed calculations and data preparations, and wrote the manuscript. All authors participated in reviewing and revising the manuscript. 

\appendix

\section{Computational details}
\label{Appendix:Comp}
\subsection{Density Functional Theory}
Density functional theory (DFT) calculations are performed within projector-augmented-wave (PAW) based Vienna {\it{Ab-initio}} Simulation Package (VASP) \cite{Kresse1996}. For all self-consistent total energy calculations and structural relaxations, plane-wave energy cutoff is set to be 550 eV. A $\Gamma$-centered 10$\times$10$\times$10 $k$-grid is used for sampling the Brillouin zone. To account for electron correlations, Ceperley-Alder local density approximation (LDA) is adopted \cite{Ceperley1980}. An additional effective Hubbard-type onsite Coulomb repulsion $U$ ($U_{\rm eff}$ = 2, 4 eV) is applied to the $d$ orbitals of Cu and Cr via a simplified rotationally-invariant DFT+$U$ approach \cite{Dudarev1998}. For relaxation of electronic degrees of freedom, energy convergence criteria is chosen to 10$^{-8}$ eV. Furthermore, spin-orbit coupling is included in the calculations to investigate topological features of the band structure. The direction of magnetization is assumed to be parallel to [001] and [111] directions in the conventional cubic unit cell.

\subsection{Tight-binding Wannier model}

As mentioned in the main text, we have employed a 36-band tight-binding (TB) model, consisting of 12 Cr-$t_{\rm 2g}$-like and 24 Se-$p$-like Wannier orbitals in the majority spin channel, for a faithful representation of the nodal line features in our model. We note in passing that, any trial to obtain TB models with fewer number of Wannier orbitals suffered from poor convergence issue that results in i) imaginary hopping integrals even in the absence of SOC or ii) disagreement in nodal line features between DFT and TB results. Because of small amount of minority spin components within $E_{\rm{F}}\pm$0.5 eV (see Fig.~\ref{fig:Fig1}(b) in the main text), we expect that this spin-up Wannier TB model is the minimal model that captures physics induced by the nodal lines of CCS.

\begin{figure}
\begin{center}
\includegraphics[origin=c,scale=0.3]{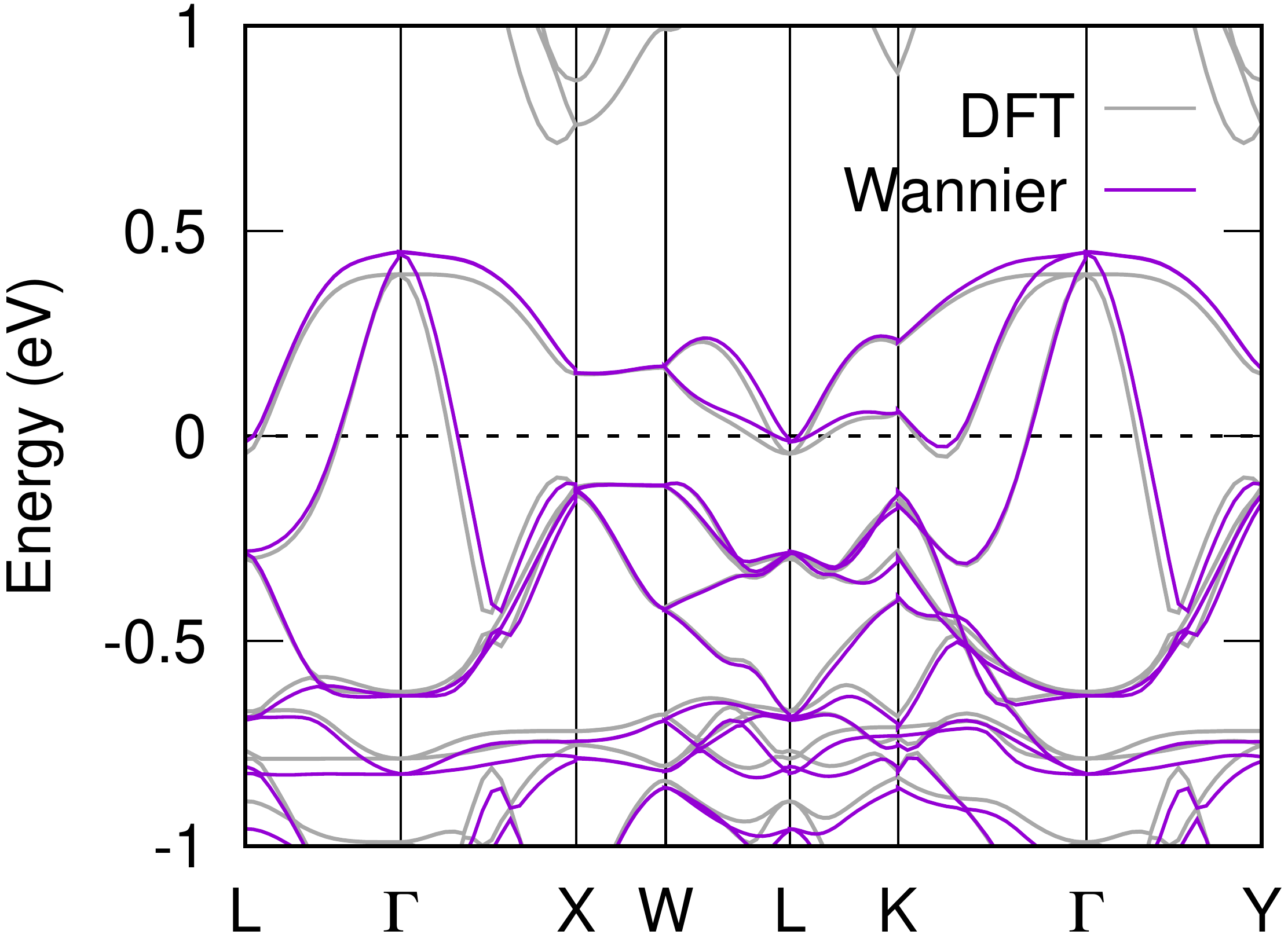}
\caption{Full DFT (grey) and Wannier interpolated (violet) band structures in the majority spin channel without spin-orbit coupling.}
\label{fig:SM_Fig1}
\end{center}
\end{figure}

Figure \ref{fig:SM_Fig1} shows band structures in the majority spin channel, obtained within full Kohn-Sham orbital basis (DFT) and minimal TB basis (Wannier). The interpolated Wannier bands are in good agreement with the full DFT bands of CCS around $E_{\rm{F}}\pm$0.5 eV, and aptly describe the key nodal-like band features on the X-W lines as shown in the figure.

Since tight-binding (TB) Hamiltonian suffices to well-describe the valence electrons within minimal basis set, here we build {\it ab-initio} TB model using Wannier function. It is well-known that most of relevant features of the topological materials originate from the relativistic effect. Hence, spin-orbit coupling is further added to the Hamiltonian. For $t_{2g}$ ($d_{yz}$, $d_{xz}$, $d_{xy}$) orbitals and magnetization direction along [001] and [111], $H_{\rm{SO}}^{t_{2g}}$ is

\begin{gather}
H_{\rm{SO}}^{001} =\lambda_{\rm{Cr}}
\begin{bmatrix*}[r]
0 & -i & 0 \\
i & 0 & \phantom{-}0 \\
0 & 0 & 0
\end{bmatrix*}
\quad
H_{\rm{SO}}^{111} =\lambda_{\rm{Cr}}
\begin{bmatrix*}[r]
0 & -i & i \\
i & 0 & -i \\
-i & \phantom{-}i & 0
\end{bmatrix*}
\end{gather}

Using T-P equivalence relation \cite{Kamimura1970}, $l(t_{2g})=-l(p)$, matrix elements of spin-orbit Hamiltonian of Se atom can be easily obtained from $H_{\rm SO}^{t_{2g}}$. In the Wannier TB model, strength of the spin-orbit coupling constants ($\lambda_{\rm{Cr}}$ = 0.02 eV \& $\lambda_{\rm{Se}}$ = 0.05 eV) are chosen such that it produces roughly same band splitting energies at the $L$ point with those from DFT calculations. 

\subsection{Chern number and anomalous Hall conductivity}

\begin{figure}
\begin{center}
\includegraphics[width=0.48\textwidth]{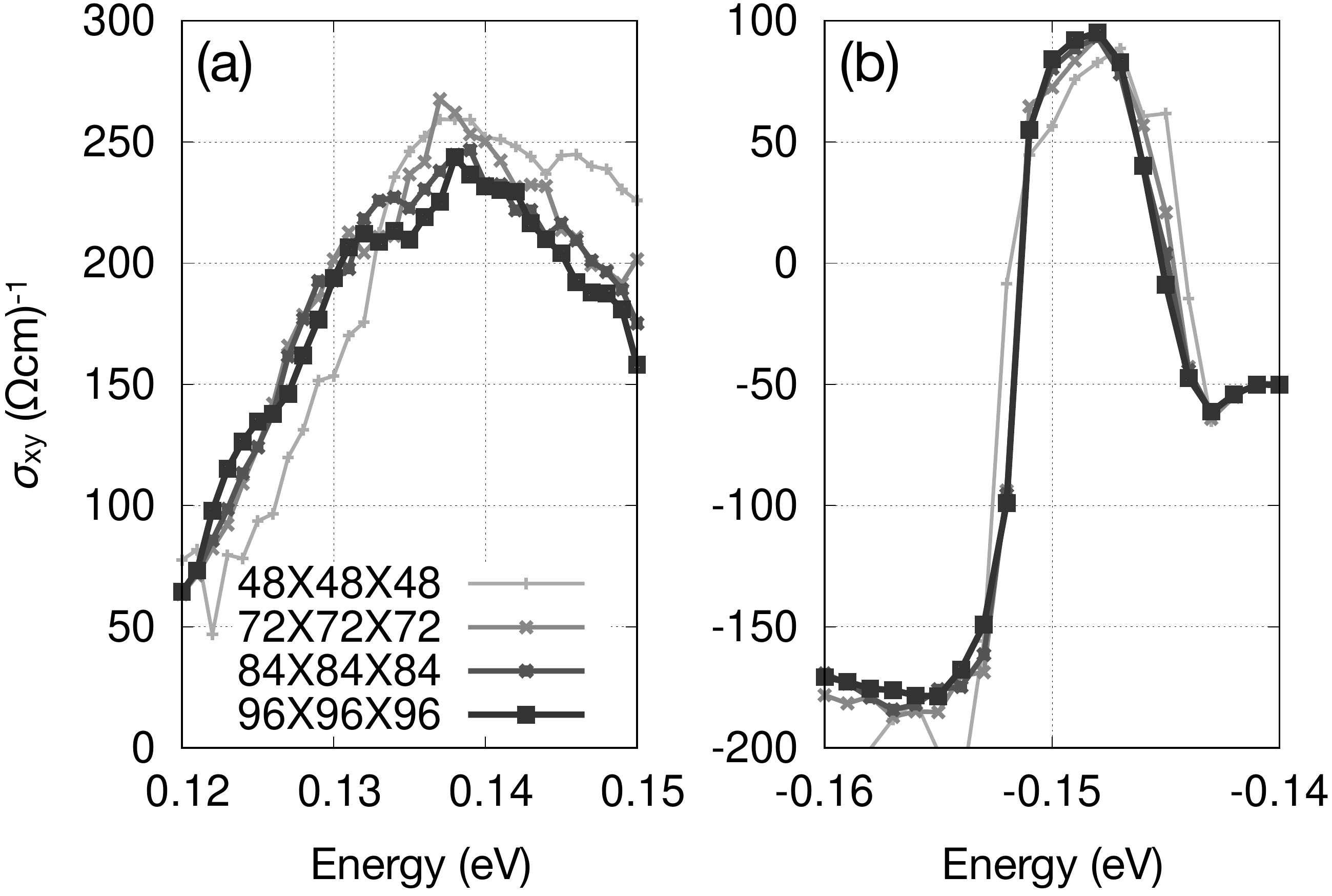}
\caption{
Computed $\sigma_{xy}(E)$ with varying $k$-grid sampling in energy ranges of (a) $0.12 \leq E \leq 0.15$ and (b) $-0.16 \leq E \leq -0.14$ eV with respect to the Fermi level. $M$ // [111] and nonzero $\lambda_{\rm Cr, Se}$ are included.
}
\label{fig:Fig8}
\end{center}
\end{figure}

To compute anomalous Hall conductivity, a numerically efficient approach, known as Fukui-Hatsugai-Suzuki (FHS) method \cite{Suzuki2005}, is employed to calculate Chern number in the discretized Brillouin zone. The FHS method computes the Berry curvature threading a $k$-plaquette defined by 4 nearest-neighboring $k$-points via the following formula,
\begin{align}
\Omega_{\alpha \beta} ({\bf k}) = \frac{1}{2\pi i} 
\ln \left[
        U_\alpha ({\bf k}) 
        U_\beta ({\bf k}+\hat{k}_\alpha)
        U_\alpha ({\bf k}+\hat{k}_\beta)^{-1}
        U_\beta ({\bf k})^{-1} 
\right],   \nonumber
\end{align}
where the link matrix $U_\alpha({\bf k})$ is defined as
\begin{align}
U_{\alpha}({\bf k}) \equiv 
\frac{{\rm det}~ \psi^\dag ({\bf k}) \psi ({\bf k} + \hat{k}_\alpha)}
{\vert {\rm det}~ \psi^\dag ({\bf k}) \psi ({\bf k} + \hat{k}_\alpha) \vert}. \nonumber
\end{align}
Here $\alpha, \beta = x,y,z$ are indices for cartesian coordinates, $\hat{k}_{\alpha}$ are unit vectors in the discretized $k$-space along the $\alpha$-direction, and $\psi({\bf k})$ is the vector of occupied states at ${\bf k}$. 

In metallic systems, the number of occupied bands can change as one crosses a Fermi surface in the $k$-space. Then the size of the link matrix $U_\alpha({\bf k})$ may not be well-defined. To circumvent this issue and apply the FHS method in metallic systems, we introduced a mixing scheme to the original formulation. When the $k$-plaquette from which a $\Omega_{\alpha \beta} ({\bf k})$ is computed intersects with a Fermi surface, then some of the vertices has different number of occupied bands compared to others. If the $k$-plaquette is small enough, there should be vertices with either $N$ or $N+1$ occupied bands only. 
In such case we compute two $\Omega^{N, N+1}_{\alpha \beta} ({\bf k})$, one with $N$-occupied bands, another with $N+1$ bands. Then we take an average of $\Omega^N_{\alpha \beta} ({\bf k})$ and $\Omega^{N+1}_{\alpha \beta} ({\bf k})$ to obtain the final $\Omega_{\alpha \beta} ({\bf k})$.

To check the validity of this trick we did a convergence test of our AHC results as a function of $k$-sampling. Fig.~\ref{fig:Fig8} presents the result of the convergence test in the vicinity of the nodal-line-induced peak of AHC close to the Fermi level, which shows a reasonable convergence at 96$\times$96$\times$96. In addition, the Berry curvature on a side face of the Brillouin zone $X$-$W$ plane is calculated with a choice of 200$\times$200 $k$-grid.

\subsection{Dynamical mean-field theory}

A combined density functional plus embedded dynamical mean-field theory (DMFT), as implemented in Rutgers EDMFTF code \cite{Haule2010,Haule2018}, is further used to study the effect of dynamical electron correlation on the nodal line in the band structure. The LDA exchange-correlation functional, which provided a more consistent result with the experimental observation, is further chosen to combine with the DMFT. Primitive Brillouin zone is sampled with a 12$\times$12$\times$12 mesh-grid. $RK_{\rm{max}}$ is set to 7. Continuous-time quantum Monte Carlo method is employed to solve quantum impurity problem in the correlated $d$-orbital subspaces of Cu and Cr. For each run, 2$\times$10$^8$ number of Monte Carlo steps are carried out. Bath temperature is set to 116 K. A simplified Ising-type density-density Coulomb interaction is used. The hybridization window is taken to be -10 to +10 eV with respect to Fermi energy. For charge self-consistent DMFT calculation, onsite Coulomb repulsion and Hund's coupling parameters are set to $U$ = 10 and $J$ = 1 eV for both Cu and Cr $d$-orbitals.

\section{Effect of static electron correlations}
\label{Appendix:HubbardU}
\begin{figure}
\begin{center}
\includegraphics[angle=-0,origin=c,scale=0.22]{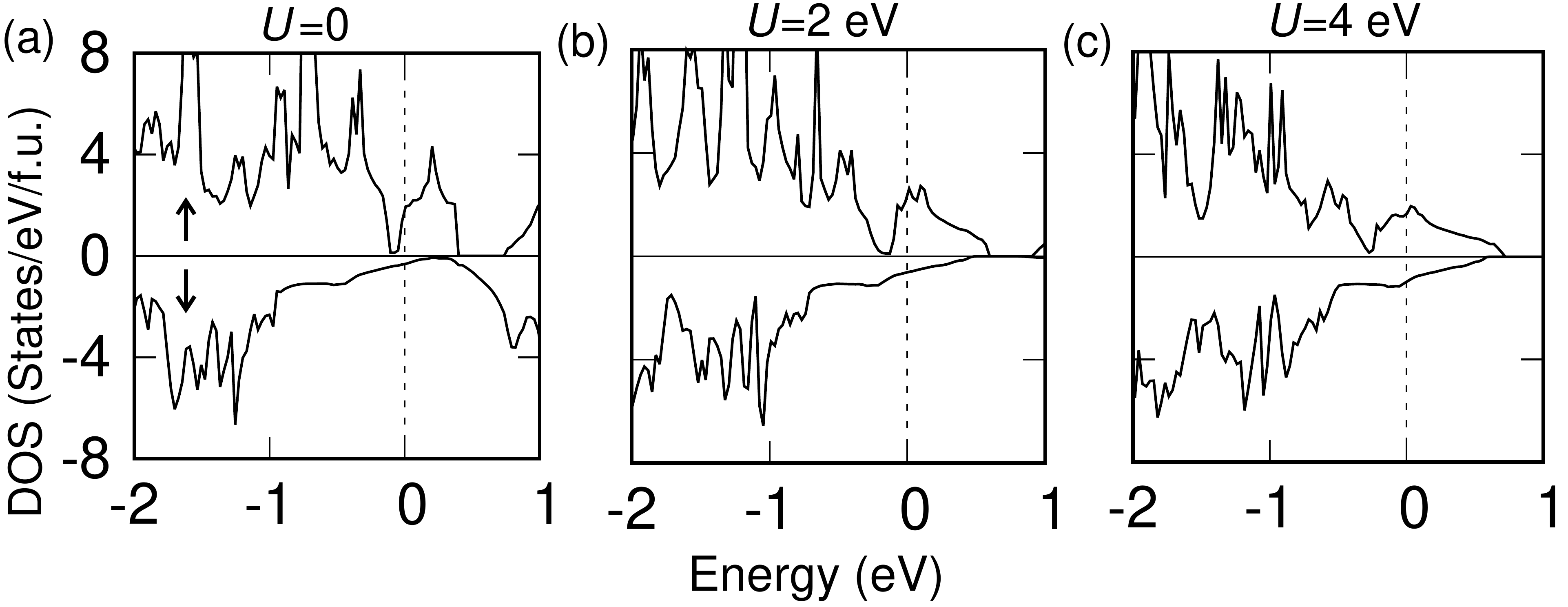}
\caption{Total density of states of CCS within (a) LDA and (b)-(c) LDA+$U$, with $U$ = 2, 4 eV demonstrate the effect of $U$ on the electronic structure.}
\label{fig:SM_Fig2}
\end{center}
\end{figure}

In general, for studies of materials having partially filled $d$ shell, often it is necessary to incorporate on-site Coulomb repulsion parameter $U$ in order to accurately calculate the electronic structure. Fig. \ref{fig:SM_Fig2} shows how the inclusion of $U$ parameter affects the ground states of CuCr$_2$Se$_4$ within the mean-field-like DFT+$U$ formalism. In the absence of $U$, states close to the Fermi-level is predominantly occupied by hybridized Cr-$t_{2g}$ - Se-$p$ states in the majority spin channel, while it is almost empty in the minority spin channel. This nearly half-metallic picture is consistent with the earlier experimental observation \cite{Kanomata2001}. When $U$ is introduced to the $d$ orbitals, density of states of minority spin channel gradually enhances and it becomes comparable to that of majority spin at the Fermi level when $U$ = 4 eV at Cr. We conclude that, in the present case, effects of DFT+$U$ correction drives the system away from half-metallic phase, so that pure LDA seems to yield better agreement with the experimental observation.

\begin{center}
\begin{figure}
\includegraphics[angle=-0,origin=c,scale=0.52]{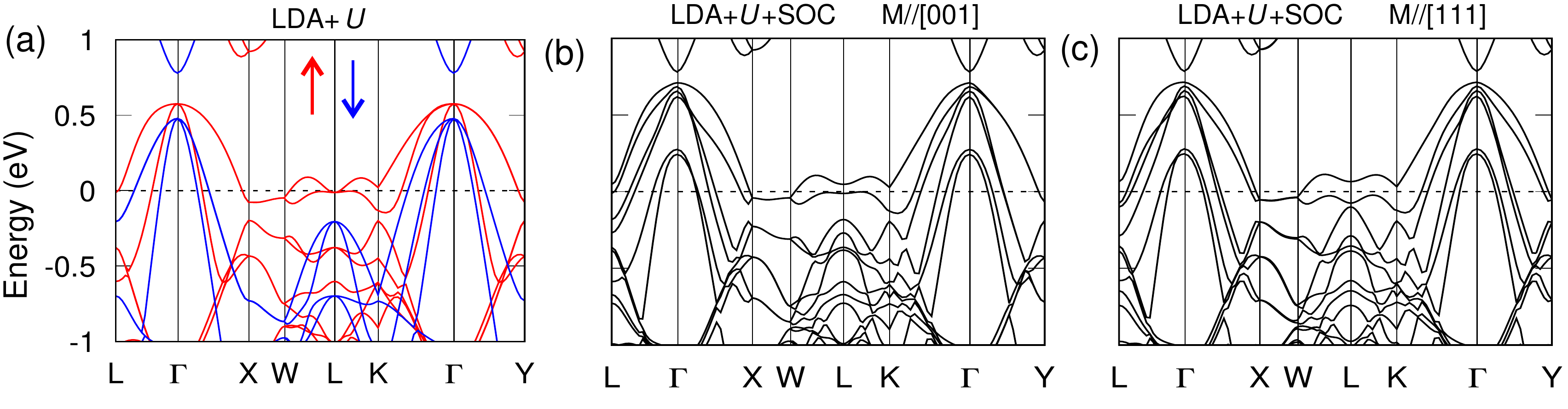}
\caption{Band structure of CCS demonstrates the effect of Hubbard $U$ (= 2 eV) and spin-orbit coupling on the electronic structure. (a) An inclusion of $U$ parameter pushes nodal line below Fermi level. (b)-(c) SOC splits the band at $L$ point.}
\label{fig:SM_Fig3}
\end{figure}
\end{center}

Figure \ref{fig:SM_Fig3} summarizes the effect of Hubbard $U$ (= 2 eV) on the electronic band structure. In the main text, it has been shown that in the LDA band structure, upper nodal line lies in the electron doped region which generates large anomalous Hall response upon electron doping. This observation is also consistent with experimental studies \cite{Cava2004}. However, band structure within DFT+$U$ shows the opposite trend; the nodal line, which were located just above the Fermi level in the absence of $U$, are pushed down in energy below the Fermi level in presence of $U$. Because the sign-changing behavior of anomalous Hall response upon electron doping is likely relevant to the nodal lines, we suspect that application of DFT+$U$ method is detrimental to the study of this compound.

\onecolumngrid
\twocolumngrid

\bibliography{cce_nodal}

\begin{thebibliography}{45}%
\makeatletter
\providecommand \@ifxundefined [1]{%
 \@ifx{#1\undefined}
}%
\providecommand \@ifnum [1]{%
 \ifnum #1\expandafter \@firstoftwo
 \else \expandafter \@secondoftwo
 \fi
}%
\providecommand \@ifx [1]{%
 \ifx #1\expandafter \@firstoftwo
 \else \expandafter \@secondoftwo
 \fi
}%
\providecommand \natexlab [1]{#1}%
\providecommand \enquote  [1]{``#1''}%
\providecommand \bibnamefont  [1]{#1}%
\providecommand \bibfnamefont [1]{#1}%
\providecommand \citenamefont [1]{#1}%
\providecommand \href@noop [0]{\@secondoftwo}%
\providecommand \href [0]{\begingroup \@sanitize@url \@href}%
\providecommand \@href[1]{\@@startlink{#1}\@@href}%
\providecommand \@@href[1]{\endgroup#1\@@endlink}%
\providecommand \@sanitize@url [0]{\catcode `\\12\catcode `\$12\catcode
  `\&12\catcode `\#12\catcode `\^12\catcode `\_12\catcode `\%12\relax}%
\providecommand \@@startlink[1]{}%
\providecommand \@@endlink[0]{}%
\providecommand \url  [0]{\begingroup\@sanitize@url \@url }%
\providecommand \@url [1]{\endgroup\@href {#1}{\urlprefix }}%
\providecommand \urlprefix  [0]{URL }%
\providecommand \Eprint [0]{\href }%
\providecommand \doibase [0]{http://dx.doi.org/}%
\providecommand \selectlanguage [0]{\@gobble}%
\providecommand \bibinfo  [0]{\@secondoftwo}%
\providecommand \bibfield  [0]{\@secondoftwo}%
\providecommand \translation [1]{[#1]}%
\providecommand \BibitemOpen [0]{}%
\providecommand \bibitemStop [0]{}%
\providecommand \bibitemNoStop [0]{.\EOS\space}%
\providecommand \EOS [0]{\spacefactor3000\relax}%
\providecommand \BibitemShut  [1]{\csname bibitem#1\endcsname}%
\let\auto@bib@innerbib\@empty
%</preamble>
\bibitem [{\citenamefont {Gilbert}(2021)}]{Gilbert2021}%
  \BibitemOpen
  \bibfield  {author} {\bibinfo {author} {\bibfnamefont {Matthew~J.}\
  \bibnamefont {Gilbert}},\ }\bibfield  {title} {\enquote {\bibinfo {title}
  {Topological electronics},}\ }\href {\doibase 10.1038/s42005-021-00569-5}
  {\bibfield  {journal} {\bibinfo  {journal} {Commun. Phys.}\ }\textbf
  {\bibinfo {volume} {4}},\ \bibinfo {pages} {1--12} (\bibinfo {year}
  {2021})}\BibitemShut {NoStop}%
\bibitem [{\citenamefont {Fan}\ and\ \citenamefont {Wang}(2016)}]{Fan2016}%
  \BibitemOpen
  \bibfield  {author} {\bibinfo {author} {\bibfnamefont {Yabin}\ \bibnamefont
  {Fan}}\ and\ \bibinfo {author} {\bibfnamefont {Kang~L.}\ \bibnamefont
  {Wang}},\ }\bibfield  {title} {\enquote {\bibinfo {title} {Spintronics based
  on topological insulators},}\ }\href {\doibase 10.1142/S2010324716400014}
  {\bibfield  {journal} {\bibinfo  {journal} {SPIN}\ }\textbf {\bibinfo
  {volume} {06}},\ \bibinfo {pages} {1640001} (\bibinfo {year}
  {2016})}\BibitemShut {NoStop}%
\bibitem [{\citenamefont {Chang}\ \emph
  {et~al.}(2015{\natexlab{a}})\citenamefont {Chang}, \citenamefont {Zhao},
  \citenamefont {Kim}, \citenamefont {Wei}, \citenamefont {Jain}, \citenamefont
  {Liu}, \citenamefont {Chan},\ and\ \citenamefont {Moodera}}]{Chang2015}%
  \BibitemOpen
  \bibfield  {author} {\bibinfo {author} {\bibfnamefont {Cui-Zu}\ \bibnamefont
  {Chang}}, \bibinfo {author} {\bibfnamefont {Weiwei}\ \bibnamefont {Zhao}},
  \bibinfo {author} {\bibfnamefont {Duk~Y.}\ \bibnamefont {Kim}}, \bibinfo
  {author} {\bibfnamefont {Peng}\ \bibnamefont {Wei}}, \bibinfo {author}
  {\bibfnamefont {J.~K.}\ \bibnamefont {Jain}}, \bibinfo {author}
  {\bibfnamefont {Chaoxing}\ \bibnamefont {Liu}}, \bibinfo {author}
  {\bibfnamefont {Moses H.~W.}\ \bibnamefont {Chan}}, \ and\ \bibinfo {author}
  {\bibfnamefont {Jagadeesh~S.}\ \bibnamefont {Moodera}},\ }\bibfield  {title}
  {\enquote {\bibinfo {title} {Zero-field dissipationless chiral edge transport
  and the nature of dissipation in the quantum anomalous hall state},}\ }\href
  {\doibase 10.1103/PhysRevLett.115.057206} {\bibfield  {journal} {\bibinfo
  {journal} {Phys. Rev. Lett.}\ }\textbf {\bibinfo {volume} {115}},\ \bibinfo
  {pages} {057206} (\bibinfo {year} {2015}{\natexlab{a}})}\BibitemShut
  {NoStop}%
\bibitem [{\citenamefont {Murakami}\ \emph {et~al.}(2003)\citenamefont
  {Murakami}, \citenamefont {Nagaosa},\ and\ \citenamefont
  {Zhang}}]{Murakami2003}%
  \BibitemOpen
  \bibfield  {author} {\bibinfo {author} {\bibfnamefont {Shuichi}\ \bibnamefont
  {Murakami}}, \bibinfo {author} {\bibfnamefont {Naoto}\ \bibnamefont
  {Nagaosa}}, \ and\ \bibinfo {author} {\bibfnamefont {Shou-Cheng}\
  \bibnamefont {Zhang}},\ }\bibfield  {title} {\enquote {\bibinfo {title}
  {Dissipationless quantum spin current at room temperature},}\ }\href
  {\doibase 10.1126/science.1087128} {\bibfield  {journal} {\bibinfo  {journal}
  {Science}\ }\textbf {\bibinfo {volume} {301}},\ \bibinfo {pages} {1348--1351}
  (\bibinfo {year} {2003})}\BibitemShut {NoStop}%
\bibitem [{\citenamefont {Culcer}\ \emph {et~al.}(2020)\citenamefont {Culcer},
  \citenamefont {Keser}, \citenamefont {Li},\ and\ \citenamefont
  {Tkachov}}]{Culcer2020}%
  \BibitemOpen
  \bibfield  {author} {\bibinfo {author} {\bibfnamefont {Dimitrie}\
  \bibnamefont {Culcer}}, \bibinfo {author} {\bibfnamefont {Ayd{\i}n~Cem}\
  \bibnamefont {Keser}}, \bibinfo {author} {\bibfnamefont {Yongqing}\
  \bibnamefont {Li}}, \ and\ \bibinfo {author} {\bibfnamefont {Grigory}\
  \bibnamefont {Tkachov}},\ }\bibfield  {title} {\enquote {\bibinfo {title}
  {Transport in two-dimensional topological materials: recent developments in
  experiment and theory},}\ }\href {\doibase 10.1088/2053-1583/ab6ff7}
  {\bibfield  {journal} {\bibinfo  {journal} {2D Mater}\ }\textbf {\bibinfo
  {volume} {7}},\ \bibinfo {pages} {022007} (\bibinfo {year}
  {2020})}\BibitemShut {NoStop}%
\bibitem [{\citenamefont {He}\ \emph {et~al.}(2018)\citenamefont {He},
  \citenamefont {Wang},\ and\ \citenamefont {Xue}}]{He2018}%
  \BibitemOpen
  \bibfield  {author} {\bibinfo {author} {\bibfnamefont {Ke}~\bibnamefont
  {He}}, \bibinfo {author} {\bibfnamefont {Yayu}\ \bibnamefont {Wang}}, \ and\
  \bibinfo {author} {\bibfnamefont {Qi-Kun}\ \bibnamefont {Xue}},\ }\bibfield
  {title} {\enquote {\bibinfo {title} {Topological materials: Quantum anomalous
  hall system},}\ }\href {\doibase 10.1146/annurev-conmatphys-033117-054144}
  {\bibfield  {journal} {\bibinfo  {journal} {Annu. Rev. Condens. Matter
  Phys.}\ }\textbf {\bibinfo {volume} {9}},\ \bibinfo {pages} {329--344}
  (\bibinfo {year} {2018})}\BibitemShut {NoStop}%
\bibitem [{\citenamefont {Tokura}\ \emph {et~al.}(2019)\citenamefont {Tokura},
  \citenamefont {Yasuda},\ and\ \citenamefont {Tsukazaki}}]{Tokura2019}%
  \BibitemOpen
  \bibfield  {author} {\bibinfo {author} {\bibfnamefont {Yoshinori}\
  \bibnamefont {Tokura}}, \bibinfo {author} {\bibfnamefont {Kenji}\
  \bibnamefont {Yasuda}}, \ and\ \bibinfo {author} {\bibfnamefont {Atsushi}\
  \bibnamefont {Tsukazaki}},\ }\bibfield  {title} {\enquote {\bibinfo {title}
  {Magnetic topological insulators},}\ }\href {\doibase
  10.1038/s42254-018-0011-5} {\bibfield  {journal} {\bibinfo  {journal} {Nat.
  Rev. Phys.}\ }\textbf {\bibinfo {volume} {1}},\ \bibinfo {pages} {126--143}
  (\bibinfo {year} {2019})}\BibitemShut {NoStop}%
\bibitem [{\citenamefont {Nagaosa}\ \emph {et~al.}(2010)\citenamefont
  {Nagaosa}, \citenamefont {Sinova}, \citenamefont {Onoda}, \citenamefont
  {MacDonald},\ and\ \citenamefont {Ong}}]{Nagaosa2010}%
  \BibitemOpen
  \bibfield  {author} {\bibinfo {author} {\bibfnamefont {Naoto}\ \bibnamefont
  {Nagaosa}}, \bibinfo {author} {\bibfnamefont {Jairo}\ \bibnamefont {Sinova}},
  \bibinfo {author} {\bibfnamefont {Shigeki}\ \bibnamefont {Onoda}}, \bibinfo
  {author} {\bibfnamefont {A.~H.}\ \bibnamefont {MacDonald}}, \ and\ \bibinfo
  {author} {\bibfnamefont {N.~P.}\ \bibnamefont {Ong}},\ }\bibfield  {title}
  {\enquote {\bibinfo {title} {Anomalous hall effect},}\ }\href {\doibase
  10.1103/RevModPhys.82.1539} {\bibfield  {journal} {\bibinfo  {journal} {Rev.
  Mod. Phys.}\ }\textbf {\bibinfo {volume} {82}},\ \bibinfo {pages}
  {1539--1592} (\bibinfo {year} {2010})}\BibitemShut {NoStop}%
\bibitem [{\citenamefont {Hasan}\ and\ \citenamefont {Kane}(2010)}]{Kane2010}%
  \BibitemOpen
  \bibfield  {author} {\bibinfo {author} {\bibfnamefont {M.~Z.}\ \bibnamefont
  {Hasan}}\ and\ \bibinfo {author} {\bibfnamefont {C.~L.}\ \bibnamefont
  {Kane}},\ }\bibfield  {title} {\enquote {\bibinfo {title} {Colloquium:
  Topological insulators},}\ }\href {\doibase 10.1103/RevModPhys.82.3045}
  {\bibfield  {journal} {\bibinfo  {journal} {Rev. Mod. Phys.}\ }\textbf
  {\bibinfo {volume} {82}},\ \bibinfo {pages} {3045--3067} (\bibinfo {year}
  {2010})}\BibitemShut {NoStop}%
\bibitem [{\citenamefont {Wan}\ \emph {et~al.}(2011)\citenamefont {Wan},
  \citenamefont {Turner}, \citenamefont {Vishwanath},\ and\ \citenamefont
  {Savrasov}}]{Vishwanath2011}%
  \BibitemOpen
  \bibfield  {author} {\bibinfo {author} {\bibfnamefont {Xiangang}\
  \bibnamefont {Wan}}, \bibinfo {author} {\bibfnamefont {Ari~M.}\ \bibnamefont
  {Turner}}, \bibinfo {author} {\bibfnamefont {Ashvin}\ \bibnamefont
  {Vishwanath}}, \ and\ \bibinfo {author} {\bibfnamefont {Sergey~Y.}\
  \bibnamefont {Savrasov}},\ }\bibfield  {title} {\enquote {\bibinfo {title}
  {Topological semimetal and fermi-arc surface states in the electronic
  structure of pyrochlore iridates},}\ }\href {\doibase
  10.1103/PhysRevB.83.205101} {\bibfield  {journal} {\bibinfo  {journal} {Phys.
  Rev. B}\ }\textbf {\bibinfo {volume} {83}},\ \bibinfo {pages} {205101}
  (\bibinfo {year} {2011})}\BibitemShut {NoStop}%
\bibitem [{\citenamefont {Yang}\ \emph {et~al.}(2011)\citenamefont {Yang},
  \citenamefont {Lu},\ and\ \citenamefont {Ran}}]{Ran2011}%
  \BibitemOpen
  \bibfield  {author} {\bibinfo {author} {\bibfnamefont {Kai-Yu}\ \bibnamefont
  {Yang}}, \bibinfo {author} {\bibfnamefont {Yuan-Ming}\ \bibnamefont {Lu}}, \
  and\ \bibinfo {author} {\bibfnamefont {Ying}\ \bibnamefont {Ran}},\
  }\bibfield  {title} {\enquote {\bibinfo {title} {Quantum hall effects in a
  weyl semimetal: Possible application in pyrochlore iridates},}\ }\href
  {\doibase 10.1103/PhysRevB.84.075129} {\bibfield  {journal} {\bibinfo
  {journal} {Phys. Rev. B}\ }\textbf {\bibinfo {volume} {84}},\ \bibinfo
  {pages} {075129} (\bibinfo {year} {2011})}\BibitemShut {NoStop}%
\bibitem [{\citenamefont {Yu}\ \emph {et~al.}(2010)\citenamefont {Yu},
  \citenamefont {Zhang}, \citenamefont {Zhang}, \citenamefont {Zhang},
  \citenamefont {Dai},\ and\ \citenamefont {Fang}}]{Yu2010}%
  \BibitemOpen
  \bibfield  {author} {\bibinfo {author} {\bibfnamefont {Rui}\ \bibnamefont
  {Yu}}, \bibinfo {author} {\bibfnamefont {Wei}\ \bibnamefont {Zhang}},
  \bibinfo {author} {\bibfnamefont {Hai-Jun}\ \bibnamefont {Zhang}}, \bibinfo
  {author} {\bibfnamefont {Shou-Cheng}\ \bibnamefont {Zhang}}, \bibinfo
  {author} {\bibfnamefont {Xi}~\bibnamefont {Dai}}, \ and\ \bibinfo {author}
  {\bibfnamefont {Zhong}\ \bibnamefont {Fang}},\ }\bibfield  {title} {\enquote
  {\bibinfo {title} {Quantized anomalous hall effect in magnetic topological
  insulators},}\ }\href {\doibase 10.1126/science.1187485} {\bibfield
  {journal} {\bibinfo  {journal} {Science}\ }\textbf {\bibinfo {volume}
  {329}},\ \bibinfo {pages} {61--64} (\bibinfo {year} {2010})}\BibitemShut
  {NoStop}%
\bibitem [{\citenamefont {Chang}\ \emph {et~al.}(2013)\citenamefont {Chang},
  \citenamefont {Zhang}, \citenamefont {Feng}, \citenamefont {Shen},
  \citenamefont {Zhang}, \citenamefont {Guo}, \citenamefont {Li}, \citenamefont
  {Ou}, \citenamefont {Wei}, \citenamefont {Wang}, \citenamefont {Ji},
  \citenamefont {Feng}, \citenamefont {Ji}, \citenamefont {Chen}, \citenamefont
  {Jia}, \citenamefont {Dai}, \citenamefont {Fang}, \citenamefont {Zhang},
  \citenamefont {He}, \citenamefont {Wang}, \citenamefont {Lu}, \citenamefont
  {Ma},\ and\ \citenamefont {Xue}}]{Chang2013}%
  \BibitemOpen
  \bibfield  {author} {\bibinfo {author} {\bibfnamefont {Cui-Zu}\ \bibnamefont
  {Chang}}, \bibinfo {author} {\bibfnamefont {Jinsong}\ \bibnamefont {Zhang}},
  \bibinfo {author} {\bibfnamefont {Xiao}\ \bibnamefont {Feng}}, \bibinfo
  {author} {\bibfnamefont {Jie}\ \bibnamefont {Shen}}, \bibinfo {author}
  {\bibfnamefont {Zuocheng}\ \bibnamefont {Zhang}}, \bibinfo {author}
  {\bibfnamefont {Minghua}\ \bibnamefont {Guo}}, \bibinfo {author}
  {\bibfnamefont {Kang}\ \bibnamefont {Li}}, \bibinfo {author} {\bibfnamefont
  {Yunbo}\ \bibnamefont {Ou}}, \bibinfo {author} {\bibfnamefont {Pang}\
  \bibnamefont {Wei}}, \bibinfo {author} {\bibfnamefont {Li-Li}\ \bibnamefont
  {Wang}}, \bibinfo {author} {\bibfnamefont {Zhong-Qing}\ \bibnamefont {Ji}},
  \bibinfo {author} {\bibfnamefont {Yang}\ \bibnamefont {Feng}}, \bibinfo
  {author} {\bibfnamefont {Shuaihua}\ \bibnamefont {Ji}}, \bibinfo {author}
  {\bibfnamefont {Xi}~\bibnamefont {Chen}}, \bibinfo {author} {\bibfnamefont
  {Jinfeng}\ \bibnamefont {Jia}}, \bibinfo {author} {\bibfnamefont
  {Xi}~\bibnamefont {Dai}}, \bibinfo {author} {\bibfnamefont {Zhong}\
  \bibnamefont {Fang}}, \bibinfo {author} {\bibfnamefont {Shou-Cheng}\
  \bibnamefont {Zhang}}, \bibinfo {author} {\bibfnamefont {Ke}~\bibnamefont
  {He}}, \bibinfo {author} {\bibfnamefont {Yayu}\ \bibnamefont {Wang}},
  \bibinfo {author} {\bibfnamefont {Li}~\bibnamefont {Lu}}, \bibinfo {author}
  {\bibfnamefont {Xu-Cun}\ \bibnamefont {Ma}}, \ and\ \bibinfo {author}
  {\bibfnamefont {Qi-Kun}\ \bibnamefont {Xue}},\ }\bibfield  {title} {\enquote
  {\bibinfo {title} {Experimental observation of the quantum anomalous hall
  effect in a magnetic topological insulator},}\ }\href {\doibase
  10.1126/science.1234414} {\bibfield  {journal} {\bibinfo  {journal}
  {Science}\ }\textbf {\bibinfo {volume} {340}},\ \bibinfo {pages} {167--170}
  (\bibinfo {year} {2013})}\BibitemShut {NoStop}%
\bibitem [{\citenamefont {Chang}\ \emph
  {et~al.}(2015{\natexlab{b}})\citenamefont {Chang}, \citenamefont {Zhao},
  \citenamefont {Kim}, \citenamefont {Zhang}, \citenamefont {Assaf},
  \citenamefont {Heiman}, \citenamefont {Zhang}, \citenamefont {Liu},
  \citenamefont {Chan},\ and\ \citenamefont {Moodera}}]{Moodera2015}%
  \BibitemOpen
  \bibfield  {author} {\bibinfo {author} {\bibfnamefont {Cui-Zu}\ \bibnamefont
  {Chang}}, \bibinfo {author} {\bibfnamefont {Weiwei}\ \bibnamefont {Zhao}},
  \bibinfo {author} {\bibfnamefont {Duk~Y.}\ \bibnamefont {Kim}}, \bibinfo
  {author} {\bibfnamefont {Haijun}\ \bibnamefont {Zhang}}, \bibinfo {author}
  {\bibfnamefont {Badih~A.}\ \bibnamefont {Assaf}}, \bibinfo {author}
  {\bibfnamefont {Don}\ \bibnamefont {Heiman}}, \bibinfo {author}
  {\bibfnamefont {Shou-Cheng}\ \bibnamefont {Zhang}}, \bibinfo {author}
  {\bibfnamefont {Chaoxing}\ \bibnamefont {Liu}}, \bibinfo {author}
  {\bibfnamefont {Moses H.~W.}\ \bibnamefont {Chan}}, \ and\ \bibinfo {author}
  {\bibfnamefont {Jagadeesh~S.}\ \bibnamefont {Moodera}},\ }\bibfield  {title}
  {\enquote {\bibinfo {title} {High-precision realization of robust quantum
  anomalous hall state in a hard ferromagnetic topological insulator},}\ }\href
  {\doibase 10.1038/nmat4204} {\bibfield  {journal} {\bibinfo  {journal} {Nat.
  Mater.}\ }\textbf {\bibinfo {volume} {14}},\ \bibinfo {pages} {473–477}
  (\bibinfo {year} {2015}{\natexlab{b}})}\BibitemShut {NoStop}%
\bibitem [{\citenamefont {Xu}\ \emph {et~al.}(2011)\citenamefont {Xu},
  \citenamefont {Weng}, \citenamefont {Wang}, \citenamefont {Dai},\ and\
  \citenamefont {Fang}}]{Zhong2011}%
  \BibitemOpen
  \bibfield  {author} {\bibinfo {author} {\bibfnamefont {Gang}\ \bibnamefont
  {Xu}}, \bibinfo {author} {\bibfnamefont {Hongming}\ \bibnamefont {Weng}},
  \bibinfo {author} {\bibfnamefont {Zhijun}\ \bibnamefont {Wang}}, \bibinfo
  {author} {\bibfnamefont {Xi}~\bibnamefont {Dai}}, \ and\ \bibinfo {author}
  {\bibfnamefont {Zhong}\ \bibnamefont {Fang}},\ }\bibfield  {title} {\enquote
  {\bibinfo {title} {Chern semimetal and the quantized anomalous hall effect in
  {H}g{C}r$_2${S}e$_4$},}\ }\href {\doibase 10.1103/PhysRevLett.107.186806}
  {\bibfield  {journal} {\bibinfo  {journal} {Phys. Rev. Lett.}\ }\textbf
  {\bibinfo {volume} {107}},\ \bibinfo {pages} {186806} (\bibinfo {year}
  {2011})}\BibitemShut {NoStop}%
\bibitem [{\citenamefont {Wang}\ \emph {et~al.}(2016)\citenamefont {Wang},
  \citenamefont {Vergniory}, \citenamefont {Kushwaha}, \citenamefont
  {Hirschberger}, \citenamefont {Chulkov}, \citenamefont {Ernst}, \citenamefont
  {Ong}, \citenamefont {Cava},\ and\ \citenamefont {Bernevig}}]{Bernevig2020}%
  \BibitemOpen
  \bibfield  {author} {\bibinfo {author} {\bibfnamefont {Zhijun}\ \bibnamefont
  {Wang}}, \bibinfo {author} {\bibfnamefont {M.~G.}\ \bibnamefont {Vergniory}},
  \bibinfo {author} {\bibfnamefont {S.}~\bibnamefont {Kushwaha}}, \bibinfo
  {author} {\bibfnamefont {Max}\ \bibnamefont {Hirschberger}}, \bibinfo
  {author} {\bibfnamefont {E.~V.}\ \bibnamefont {Chulkov}}, \bibinfo {author}
  {\bibfnamefont {A.}~\bibnamefont {Ernst}}, \bibinfo {author} {\bibfnamefont
  {N.~P.}\ \bibnamefont {Ong}}, \bibinfo {author} {\bibfnamefont {Robert~J.}\
  \bibnamefont {Cava}}, \ and\ \bibinfo {author} {\bibfnamefont {B.~Andrei}\
  \bibnamefont {Bernevig}},\ }\bibfield  {title} {\enquote {\bibinfo {title}
  {Time-reversal-breaking weyl fermions in magnetic heusler alloys},}\ }\href
  {\doibase 10.1103/PhysRevLett.117.236401} {\bibfield  {journal} {\bibinfo
  {journal} {Phys. Rev. Lett.}\ }\textbf {\bibinfo {volume} {117}},\ \bibinfo
  {pages} {236401} (\bibinfo {year} {2016})}\BibitemShut {NoStop}%
\bibitem [{\citenamefont {Nakatsuji}\ \emph {et~al.}(2015)\citenamefont
  {Nakatsuji}, \citenamefont {Kiyohara},\ and\ \citenamefont
  {Higo}}]{Higo2015}%
  \BibitemOpen
  \bibfield  {author} {\bibinfo {author} {\bibfnamefont {Satoru}\ \bibnamefont
  {Nakatsuji}}, \bibinfo {author} {\bibfnamefont {Naoki}\ \bibnamefont
  {Kiyohara}}, \ and\ \bibinfo {author} {\bibfnamefont {Tomoya}\ \bibnamefont
  {Higo}},\ }\bibfield  {title} {\enquote {\bibinfo {title} {Large anomalous
  hall effect in a non-collinear antiferromagnet at room temperature},}\ }\href
  {\doibase 10.1038/nature15723} {\bibfield  {journal} {\bibinfo  {journal}
  {Nature}\ }\textbf {\bibinfo {volume} {527}},\ \bibinfo {pages} {212--215}
  (\bibinfo {year} {2015})}\BibitemShut {NoStop}%
\bibitem [{\citenamefont {Kim}\ \emph {et~al.}(2018)\citenamefont {Kim},
  \citenamefont {Seo}, \citenamefont {Lee}, \citenamefont {Ko}, \citenamefont
  {Kim}, \citenamefont {Jang}, \citenamefont {Ok}, \citenamefont {Lee},
  \citenamefont {Jo}, \citenamefont {Kang}, \citenamefont {Shim}, \citenamefont
  {Kim}, \citenamefont {Yeom}, \citenamefont {II~Min}, \citenamefont {Yang},\
  and\ \citenamefont {Kim}}]{Kim2018}%
  \BibitemOpen
  \bibfield  {author} {\bibinfo {author} {\bibfnamefont {Kyoo}\ \bibnamefont
  {Kim}}, \bibinfo {author} {\bibfnamefont {Junho}\ \bibnamefont {Seo}},
  \bibinfo {author} {\bibfnamefont {Eunwoo}\ \bibnamefont {Lee}}, \bibinfo
  {author} {\bibfnamefont {K.-T.}\ \bibnamefont {Ko}}, \bibinfo {author}
  {\bibfnamefont {B.~S.}\ \bibnamefont {Kim}}, \bibinfo {author} {\bibfnamefont
  {Bo~Gyu}\ \bibnamefont {Jang}}, \bibinfo {author} {\bibfnamefont {Jong~Mok}\
  \bibnamefont {Ok}}, \bibinfo {author} {\bibfnamefont {Jinwon}\ \bibnamefont
  {Lee}}, \bibinfo {author} {\bibfnamefont {Youn~Jung}\ \bibnamefont {Jo}},
  \bibinfo {author} {\bibfnamefont {Woun}\ \bibnamefont {Kang}}, \bibinfo
  {author} {\bibfnamefont {Ji~Hoon}\ \bibnamefont {Shim}}, \bibinfo {author}
  {\bibfnamefont {C.}~\bibnamefont {Kim}}, \bibinfo {author} {\bibfnamefont
  {Han~Woong}\ \bibnamefont {Yeom}}, \bibinfo {author} {\bibfnamefont {Byung}\
  \bibnamefont {II~Min}}, \bibinfo {author} {\bibfnamefont {Bohm-Jung}\
  \bibnamefont {Yang}}, \ and\ \bibinfo {author} {\bibfnamefont {Jun~Sung}\
  \bibnamefont {Kim}},\ }\bibfield  {title} {\enquote {\bibinfo {title} {Large
  anomalous hall current induced by topological nodal lines in a ferromagnetic
  van der waals semimetal},}\ }\href {\doibase 10.1038/s41563-018-0132-3}
  {\bibfield  {journal} {\bibinfo  {journal} {Nat Mater}\ }\textbf {\bibinfo
  {volume} {17}},\ \bibinfo {pages} {794--799} (\bibinfo {year}
  {2018})}\BibitemShut {NoStop}%
\bibitem [{\citenamefont {Yang}\ \emph {et~al.}(2017)\citenamefont {Yang},
  \citenamefont {Bojesen}, \citenamefont {Morimoto},\ and\ \citenamefont
  {Furusaki}}]{Yang2017}%
  \BibitemOpen
  \bibfield  {author} {\bibinfo {author} {\bibfnamefont {Bohm-Jung}\
  \bibnamefont {Yang}}, \bibinfo {author} {\bibfnamefont {Troels~Arnfred}\
  \bibnamefont {Bojesen}}, \bibinfo {author} {\bibfnamefont {Takahiro}\
  \bibnamefont {Morimoto}}, \ and\ \bibinfo {author} {\bibfnamefont {Akira}\
  \bibnamefont {Furusaki}},\ }\bibfield  {title} {\enquote {\bibinfo {title}
  {Topological semimetals protected by off-centered symmetries in nonsymmorphic
  crystals},}\ }\href {\doibase 10.1103/PhysRevB.95.075135} {\bibfield
  {journal} {\bibinfo  {journal} {Phys. Rev. B}\ }\textbf {\bibinfo {volume}
  {95}},\ \bibinfo {pages} {075135} (\bibinfo {year} {2017})}\BibitemShut
  {NoStop}%
\bibitem [{\citenamefont {Sun}\ \emph {et~al.}(2017)\citenamefont {Sun},
  \citenamefont {Zhang}, \citenamefont {Liu}, \citenamefont {Felser},\ and\
  \citenamefont {Yan}}]{Sun2017}%
  \BibitemOpen
  \bibfield  {author} {\bibinfo {author} {\bibfnamefont {Yan}\ \bibnamefont
  {Sun}}, \bibinfo {author} {\bibfnamefont {Yang}\ \bibnamefont {Zhang}},
  \bibinfo {author} {\bibfnamefont {Chao-Xing}\ \bibnamefont {Liu}}, \bibinfo
  {author} {\bibfnamefont {Claudia}\ \bibnamefont {Felser}}, \ and\ \bibinfo
  {author} {\bibfnamefont {Binghai}\ \bibnamefont {Yan}},\ }\bibfield  {title}
  {\enquote {\bibinfo {title} {Dirac nodal lines and induced spin hall effect
  in metallic rutile oxides},}\ }\href {\doibase 10.1103/PhysRevB.95.235104}
  {\bibfield  {journal} {\bibinfo  {journal} {Phys. Rev. B}\ }\textbf {\bibinfo
  {volume} {95}},\ \bibinfo {pages} {235104} (\bibinfo {year}
  {2017})}\BibitemShut {NoStop}%
\bibitem [{\citenamefont {Yang}\ \emph {et~al.}(2018)\citenamefont {Yang},
  \citenamefont {Yang}, \citenamefont {Derunova}, \citenamefont {Parkin},
  \citenamefont {Yan},\ and\ \citenamefont {Ali}}]{Yang2018}%
  \BibitemOpen
  \bibfield  {author} {\bibinfo {author} {\bibfnamefont {Shuo-Ying}\
  \bibnamefont {Yang}}, \bibinfo {author} {\bibfnamefont {Hao}\ \bibnamefont
  {Yang}}, \bibinfo {author} {\bibfnamefont {Elena}\ \bibnamefont {Derunova}},
  \bibinfo {author} {\bibfnamefont {Stuart S.~P.}\ \bibnamefont {Parkin}},
  \bibinfo {author} {\bibfnamefont {Binghai}\ \bibnamefont {Yan}}, \ and\
  \bibinfo {author} {\bibfnamefont {Mazhar~N.}\ \bibnamefont {Ali}},\
  }\bibfield  {title} {\enquote {\bibinfo {title} {Symmetry demanded
  topological nodal-line materials},}\ }\href {\doibase
  10.1080/23746149.2017.1414631} {\bibfield  {journal} {\bibinfo  {journal}
  {Advances in Physics: X}\ }\textbf {\bibinfo {volume} {3}},\ \bibinfo {pages}
  {1414631} (\bibinfo {year} {2018})},\ \Eprint
  {http://arxiv.org/abs/https://doi.org/10.1080/23746149.2017.1414631}
  {https://doi.org/10.1080/23746149.2017.1414631} \BibitemShut {NoStop}%
\bibitem [{\citenamefont {Lee}\ \emph {et~al.}(2004{\natexlab{a}})\citenamefont
  {Lee}, \citenamefont {Watauchi}, \citenamefont {Miller}, \citenamefont
  {Cava},\ and\ \citenamefont {Ong}}]{Ong2004}%
  \BibitemOpen
  \bibfield  {author} {\bibinfo {author} {\bibfnamefont {Wei-Li}\ \bibnamefont
  {Lee}}, \bibinfo {author} {\bibfnamefont {Satoshi}\ \bibnamefont {Watauchi}},
  \bibinfo {author} {\bibfnamefont {V.~L.}\ \bibnamefont {Miller}}, \bibinfo
  {author} {\bibfnamefont {R.~J.}\ \bibnamefont {Cava}}, \ and\ \bibinfo
  {author} {\bibfnamefont {N.~P.}\ \bibnamefont {Ong}},\ }\bibfield  {title}
  {\enquote {\bibinfo {title} {Dissipationless anomalous hall current in the
  ferromagnetic spinel {C}u{C}r$_2${S}e$_{4-x}${B}r$_x$},}\ }\href {\doibase
  10.1126/science.1094383} {\bibfield  {journal} {\bibinfo  {journal}
  {Science}\ }\textbf {\bibinfo {volume} {303}},\ \bibinfo {pages} {1647--1649}
  (\bibinfo {year} {2004}{\natexlab{a}})}\BibitemShut {NoStop}%
\bibitem [{\citenamefont {Lee}\ \emph {et~al.}(2004{\natexlab{b}})\citenamefont
  {Lee}, \citenamefont {Watauchi}, \citenamefont {Miller}, \citenamefont
  {Cava},\ and\ \citenamefont {Ong}}]{Cava2004}%
  \BibitemOpen
  \bibfield  {author} {\bibinfo {author} {\bibfnamefont {Wei-Li}\ \bibnamefont
  {Lee}}, \bibinfo {author} {\bibfnamefont {S.}~\bibnamefont {Watauchi}},
  \bibinfo {author} {\bibfnamefont {V.~L.}\ \bibnamefont {Miller}}, \bibinfo
  {author} {\bibfnamefont {R.~J.}\ \bibnamefont {Cava}}, \ and\ \bibinfo
  {author} {\bibfnamefont {N.~P.}\ \bibnamefont {Ong}},\ }\bibfield  {title}
  {\enquote {\bibinfo {title} {Anomalous hall heat current and nernst effect in
  the {C}u{C}r$_2${S}e$_{4-x}${B}r$_x$ ferromagnet},}\ }\href {\doibase
  10.1103/PhysRevLett.93.226601} {\bibfield  {journal} {\bibinfo  {journal}
  {Phys. Rev. Lett.}\ }\textbf {\bibinfo {volume} {93}},\ \bibinfo {pages}
  {226601} (\bibinfo {year} {2004}{\natexlab{b}})}\BibitemShut {NoStop}%
\bibitem [{\citenamefont {Xiao}\ \emph {et~al.}(2006)\citenamefont {Xiao},
  \citenamefont {Yao}, \citenamefont {Fang},\ and\ \citenamefont
  {Niu}}]{Niu2006}%
  \BibitemOpen
  \bibfield  {author} {\bibinfo {author} {\bibfnamefont {Di}~\bibnamefont
  {Xiao}}, \bibinfo {author} {\bibfnamefont {Yugui}\ \bibnamefont {Yao}},
  \bibinfo {author} {\bibfnamefont {Zhong}\ \bibnamefont {Fang}}, \ and\
  \bibinfo {author} {\bibfnamefont {Qian}\ \bibnamefont {Niu}},\ }\bibfield
  {title} {\enquote {\bibinfo {title} {Berry-phase effect in anomalous
  thermoelectric transport},}\ }\href {\doibase 10.1103/PhysRevLett.97.026603}
  {\bibfield  {journal} {\bibinfo  {journal} {Phys. Rev. Lett.}\ }\textbf
  {\bibinfo {volume} {97}},\ \bibinfo {pages} {026603} (\bibinfo {year}
  {2006})}\BibitemShut {NoStop}%
\bibitem [{\citenamefont {Yao}\ \emph {et~al.}(2007)\citenamefont {Yao},
  \citenamefont {Liang}, \citenamefont {Xiao}, \citenamefont {Niu},
  \citenamefont {Shen}, \citenamefont {Dai},\ and\ \citenamefont
  {Fang}}]{Zhong2007}%
  \BibitemOpen
  \bibfield  {author} {\bibinfo {author} {\bibfnamefont {Yugui}\ \bibnamefont
  {Yao}}, \bibinfo {author} {\bibfnamefont {Yongcheng}\ \bibnamefont {Liang}},
  \bibinfo {author} {\bibfnamefont {Di}~\bibnamefont {Xiao}}, \bibinfo {author}
  {\bibfnamefont {Qian}\ \bibnamefont {Niu}}, \bibinfo {author} {\bibfnamefont
  {Shun-Qing}\ \bibnamefont {Shen}}, \bibinfo {author} {\bibfnamefont
  {X.}~\bibnamefont {Dai}}, \ and\ \bibinfo {author} {\bibfnamefont {Zhong}\
  \bibnamefont {Fang}},\ }\bibfield  {title} {\enquote {\bibinfo {title}
  {Theoretical evidence of the berry-phase mechanism in anomalous hall
  transport: First-principles studies of {C}u{C}r$_2${S}e$_{4-x}${B}r$_x$},}\
  }\href {\doibase 10.1103/PhysRevB.75.020401} {\bibfield  {journal} {\bibinfo
  {journal} {Phys. Rev. B}\ }\textbf {\bibinfo {volume} {75}},\ \bibinfo
  {pages} {020401} (\bibinfo {year} {2007})}\BibitemShut {NoStop}%
\bibitem [{\citenamefont {Nakatani}\ \emph {et~al.}(1978)\citenamefont
  {Nakatani}, \citenamefont {Nose},\ and\ \citenamefont
  {Masumoto}}]{Masumoto1978}%
  \BibitemOpen
  \bibfield  {author} {\bibinfo {author} {\bibfnamefont {I.}~\bibnamefont
  {Nakatani}}, \bibinfo {author} {\bibfnamefont {H.}~\bibnamefont {Nose}}, \
  and\ \bibinfo {author} {\bibfnamefont {K.}~\bibnamefont {Masumoto}},\
  }\bibfield  {title} {\enquote {\bibinfo {title} {Magnetic properties of
  cucr2se4 single crystals},}\ }\href {\doibase 10.1016/0022-3697(78)90008-2}
  {\bibfield  {journal} {\bibinfo  {journal} {J. Phys. Chem. Solids}\ }\textbf
  {\bibinfo {volume} {39}},\ \bibinfo {pages} {743--749} (\bibinfo {year}
  {1978})}\BibitemShut {NoStop}%
\bibitem [{\citenamefont {Kresse}\ and\ \citenamefont
  {Furthm\"{u}ller}(1996)}]{Kresse1996}%
  \BibitemOpen
  \bibfield  {author} {\bibinfo {author} {\bibfnamefont {G.}~\bibnamefont
  {Kresse}}\ and\ \bibinfo {author} {\bibfnamefont {J.}~\bibnamefont
  {Furthm\"{u}ller}},\ }\bibfield  {title} {\enquote {\bibinfo {title}
  {Efficient iterative schemes for ab initio total-energy calculations using a
  plane-wave basis set},}\ }\href {\doibase 10.1103/PhysRevB.54.11169}
  {\bibfield  {journal} {\bibinfo  {journal} {Phys. Rev. B}\ }\textbf {\bibinfo
  {volume} {54}},\ \bibinfo {pages} {11169--11186} (\bibinfo {year}
  {1996})}\BibitemShut {NoStop}%
\bibitem [{\citenamefont {Souza}\ \emph {et~al.}(2001)\citenamefont {Souza},
  \citenamefont {Marzari},\ and\ \citenamefont {Vanderbilt}}]{Souza2001}%
  \BibitemOpen
  \bibfield  {author} {\bibinfo {author} {\bibfnamefont {Ivo}\ \bibnamefont
  {Souza}}, \bibinfo {author} {\bibfnamefont {Nicola}\ \bibnamefont {Marzari}},
  \ and\ \bibinfo {author} {\bibfnamefont {David}\ \bibnamefont {Vanderbilt}},\
  }\bibfield  {title} {\enquote {\bibinfo {title} {Maximally localized wannier
  functions for entangled energy bands},}\ }\href {\doibase
  10.1103/PhysRevB.65.035109} {\bibfield  {journal} {\bibinfo  {journal} {Phys.
  Rev. B}\ }\textbf {\bibinfo {volume} {65}},\ \bibinfo {pages} {035109}
  (\bibinfo {year} {2001})}\BibitemShut {NoStop}%
\bibitem [{\citenamefont {Pizzi}\ \emph {et~al.}(2020)\citenamefont {Pizzi},
  \citenamefont {Vitale}, \citenamefont {Arita}, \citenamefont {Bl{\"{e}}gel},
  \citenamefont {Freimuth}, \citenamefont {G{\'{e}}ranton}, \citenamefont
  {Gibertini}, \citenamefont {Gresch}, \citenamefont {Johnson}, \citenamefont
  {Koretsune}, \citenamefont {Iba{\~{n}}ez-Azpiroz}, \citenamefont {Lee},
  \citenamefont {Lihm}, \citenamefont {Marchand}, \citenamefont {Marrazzo},
  \citenamefont {Mokrousov}, \citenamefont {Mustafa}, \citenamefont {Nohara},
  \citenamefont {Nomura}, \citenamefont {Paulatto}, \citenamefont
  {Ponc{\'{e}}}, \citenamefont {Ponweiser}, \citenamefont {Qiao}, \citenamefont
  {Thöle}, \citenamefont {Tsirkin}, \citenamefont {Wierzbowska}, \citenamefont
  {Marzari}, \citenamefont {Vanderbilt}, \citenamefont {Souza}, \citenamefont
  {Mostofi},\ and\ \citenamefont {Yates}}]{Pizzi2020}%
  \BibitemOpen
  \bibfield  {author} {\bibinfo {author} {\bibfnamefont {Giovanni}\
  \bibnamefont {Pizzi}}, \bibinfo {author} {\bibfnamefont {Valerio}\
  \bibnamefont {Vitale}}, \bibinfo {author} {\bibfnamefont {Ryotaro}\
  \bibnamefont {Arita}}, \bibinfo {author} {\bibfnamefont {Stefan}\
  \bibnamefont {Bl{\"{e}}gel}}, \bibinfo {author} {\bibfnamefont {Frank}\
  \bibnamefont {Freimuth}}, \bibinfo {author} {\bibfnamefont {Guillaume}\
  \bibnamefont {G{\'{e}}ranton}}, \bibinfo {author} {\bibfnamefont {Marco}\
  \bibnamefont {Gibertini}}, \bibinfo {author} {\bibfnamefont {Dominik}\
  \bibnamefont {Gresch}}, \bibinfo {author} {\bibfnamefont {Charles}\
  \bibnamefont {Johnson}}, \bibinfo {author} {\bibfnamefont {Takashi}\
  \bibnamefont {Koretsune}}, \bibinfo {author} {\bibfnamefont {Julen}\
  \bibnamefont {Iba{\~{n}}ez-Azpiroz}}, \bibinfo {author} {\bibfnamefont
  {Hyungjun}\ \bibnamefont {Lee}}, \bibinfo {author} {\bibfnamefont {Jae-Mo}\
  \bibnamefont {Lihm}}, \bibinfo {author} {\bibfnamefont {Daniel}\ \bibnamefont
  {Marchand}}, \bibinfo {author} {\bibfnamefont {Antimo}\ \bibnamefont
  {Marrazzo}}, \bibinfo {author} {\bibfnamefont {Yuriy}\ \bibnamefont
  {Mokrousov}}, \bibinfo {author} {\bibfnamefont {Jamal~I}\ \bibnamefont
  {Mustafa}}, \bibinfo {author} {\bibfnamefont {Yoshiro}\ \bibnamefont
  {Nohara}}, \bibinfo {author} {\bibfnamefont {Yusuke}\ \bibnamefont {Nomura}},
  \bibinfo {author} {\bibfnamefont {Lorenzo}\ \bibnamefont {Paulatto}},
  \bibinfo {author} {\bibfnamefont {Samuel}\ \bibnamefont {Ponc{\'{e}}}},
  \bibinfo {author} {\bibfnamefont {Thomas}\ \bibnamefont {Ponweiser}},
  \bibinfo {author} {\bibfnamefont {Junfeng}\ \bibnamefont {Qiao}}, \bibinfo
  {author} {\bibfnamefont {Florian}\ \bibnamefont {Thöle}}, \bibinfo {author}
  {\bibfnamefont {Stepan~S}\ \bibnamefont {Tsirkin}}, \bibinfo {author}
  {\bibfnamefont {Ma{\l}gorzata}\ \bibnamefont {Wierzbowska}}, \bibinfo
  {author} {\bibfnamefont {Nicola}\ \bibnamefont {Marzari}}, \bibinfo {author}
  {\bibfnamefont {David}\ \bibnamefont {Vanderbilt}}, \bibinfo {author}
  {\bibfnamefont {Ivo}\ \bibnamefont {Souza}}, \bibinfo {author} {\bibfnamefont
  {Arash~A}\ \bibnamefont {Mostofi}}, \ and\ \bibinfo {author} {\bibfnamefont
  {Jonathan~R}\ \bibnamefont {Yates}},\ }\bibfield  {title} {\enquote {\bibinfo
  {title} {Wannier90 as a community code: new features and applications},}\
  }\href {\doibase 10.1088/1361-648x/ab51ff} {\bibfield  {journal} {\bibinfo
  {journal} {J. Phys.: Condens Matter}\ }\textbf {\bibinfo {volume} {32}},\
  \bibinfo {pages} {165902} (\bibinfo {year} {2020})}\BibitemShut {NoStop}%
\bibitem [{\citenamefont {Kimura}\ \emph {et~al.}(2001)\citenamefont {Kimura},
  \citenamefont {Matsuno}, \citenamefont {Okabayashi}, \citenamefont
  {Fujimori}, \citenamefont {Shishidou}, \citenamefont {Kulatov},\ and\
  \citenamefont {Kanomata}}]{Kanomata2001}%
  \BibitemOpen
  \bibfield  {author} {\bibinfo {author} {\bibfnamefont {A.}~\bibnamefont
  {Kimura}}, \bibinfo {author} {\bibfnamefont {J.}~\bibnamefont {Matsuno}},
  \bibinfo {author} {\bibfnamefont {J.}~\bibnamefont {Okabayashi}}, \bibinfo
  {author} {\bibfnamefont {A.}~\bibnamefont {Fujimori}}, \bibinfo {author}
  {\bibfnamefont {T.}~\bibnamefont {Shishidou}}, \bibinfo {author}
  {\bibfnamefont {E.}~\bibnamefont {Kulatov}}, \ and\ \bibinfo {author}
  {\bibfnamefont {T.}~\bibnamefont {Kanomata}},\ }\bibfield  {title} {\enquote
  {\bibinfo {title} {Soft x-ray magnetic circular dichroism study of the
  ferromagnetic spinel-type cr chalcogenides},}\ }\href {\doibase
  10.1103/PhysRevB.63.224420} {\bibfield  {journal} {\bibinfo  {journal} {Phys.
  Rev. B}\ }\textbf {\bibinfo {volume} {63}},\ \bibinfo {pages} {224420}
  (\bibinfo {year} {2001})}\BibitemShut {NoStop}%
\bibitem [{\citenamefont {Feng}\ \emph {et~al.}(2019)\citenamefont {Feng},
  \citenamefont {Zhang}, \citenamefont {Feng}, \citenamefont {Fu},
  \citenamefont {Wu}, \citenamefont {Miyamoto}, \citenamefont {He},
  \citenamefont {Chen}, \citenamefont {Wu}, \citenamefont {Shimada},
  \citenamefont {Okuda},\ and\ \citenamefont {Yao}}]{Yugui2019}%
  \BibitemOpen
  \bibfield  {author} {\bibinfo {author} {\bibfnamefont {Baojie}\ \bibnamefont
  {Feng}}, \bibinfo {author} {\bibfnamefont {Run-Wu}\ \bibnamefont {Zhang}},
  \bibinfo {author} {\bibfnamefont {Ya}~\bibnamefont {Feng}}, \bibinfo {author}
  {\bibfnamefont {Botao}\ \bibnamefont {Fu}}, \bibinfo {author} {\bibfnamefont
  {Shilong}\ \bibnamefont {Wu}}, \bibinfo {author} {\bibfnamefont {Koji}\
  \bibnamefont {Miyamoto}}, \bibinfo {author} {\bibfnamefont {Shaolong}\
  \bibnamefont {He}}, \bibinfo {author} {\bibfnamefont {Lan}\ \bibnamefont
  {Chen}}, \bibinfo {author} {\bibfnamefont {Kehui}\ \bibnamefont {Wu}},
  \bibinfo {author} {\bibfnamefont {Kenya}\ \bibnamefont {Shimada}}, \bibinfo
  {author} {\bibfnamefont {Taichi}\ \bibnamefont {Okuda}}, \ and\ \bibinfo
  {author} {\bibfnamefont {Yugui}\ \bibnamefont {Yao}},\ }\bibfield  {title}
  {\enquote {\bibinfo {title} {Discovery of weyl nodal lines in a single-layer
  ferromagnet},}\ }\href {\doibase 10.1103/PhysRevLett.123.116401} {\bibfield
  {journal} {\bibinfo  {journal} {Phys. Rev. Lett.}\ }\textbf {\bibinfo
  {volume} {123}},\ \bibinfo {pages} {116401} (\bibinfo {year}
  {2019})}\BibitemShut {NoStop}%
\bibitem [{\citenamefont {Kim}\ \emph {et~al.}(2015)\citenamefont {Kim},
  \citenamefont {Wieder}, \citenamefont {Kane},\ and\ \citenamefont
  {Rappe}}]{Kane2015}%
  \BibitemOpen
  \bibfield  {author} {\bibinfo {author} {\bibfnamefont {Youngkuk}\
  \bibnamefont {Kim}}, \bibinfo {author} {\bibfnamefont {Benjamin~J.}\
  \bibnamefont {Wieder}}, \bibinfo {author} {\bibfnamefont {C.~L.}\
  \bibnamefont {Kane}}, \ and\ \bibinfo {author} {\bibfnamefont {Andrew~M.}\
  \bibnamefont {Rappe}},\ }\bibfield  {title} {\enquote {\bibinfo {title}
  {Dirac line nodes in inversion-symmetric crystals},}\ }\href {\doibase
  10.1103/PhysRevLett.115.036806} {\bibfield  {journal} {\bibinfo  {journal}
  {Phys. Rev. Lett.}\ }\textbf {\bibinfo {volume} {115}},\ \bibinfo {pages}
  {036806} (\bibinfo {year} {2015})}\BibitemShut {NoStop}%
\bibitem [{\citenamefont {Manna}\ \emph {et~al.}(2018)\citenamefont {Manna},
  \citenamefont {Muechler}, \citenamefont {Kao}, \citenamefont {Stinshoff},
  \citenamefont {Zhang}, \citenamefont {Gooth}, \citenamefont {Kumar},
  \citenamefont {Kreiner}, \citenamefont {Koepernik}, \citenamefont {Car},
  \citenamefont {K\"ubler}, \citenamefont {Fecher}, \citenamefont {Shekhar},
  \citenamefont {Sun},\ and\ \citenamefont {Felser}}]{Manna2018}%
  \BibitemOpen
  \bibfield  {author} {\bibinfo {author} {\bibfnamefont {Kaustuv}\ \bibnamefont
  {Manna}}, \bibinfo {author} {\bibfnamefont {Lukas}\ \bibnamefont {Muechler}},
  \bibinfo {author} {\bibfnamefont {Ting-Hui}\ \bibnamefont {Kao}}, \bibinfo
  {author} {\bibfnamefont {Rolf}\ \bibnamefont {Stinshoff}}, \bibinfo {author}
  {\bibfnamefont {Yang}\ \bibnamefont {Zhang}}, \bibinfo {author}
  {\bibfnamefont {Johannes}\ \bibnamefont {Gooth}}, \bibinfo {author}
  {\bibfnamefont {Nitesh}\ \bibnamefont {Kumar}}, \bibinfo {author}
  {\bibfnamefont {Guido}\ \bibnamefont {Kreiner}}, \bibinfo {author}
  {\bibfnamefont {Klaus}\ \bibnamefont {Koepernik}}, \bibinfo {author}
  {\bibfnamefont {Roberto}\ \bibnamefont {Car}}, \bibinfo {author}
  {\bibfnamefont {J\"urgen}\ \bibnamefont {K\"ubler}}, \bibinfo {author}
  {\bibfnamefont {Gerhard~H.}\ \bibnamefont {Fecher}}, \bibinfo {author}
  {\bibfnamefont {Chandra}\ \bibnamefont {Shekhar}}, \bibinfo {author}
  {\bibfnamefont {Yan}\ \bibnamefont {Sun}}, \ and\ \bibinfo {author}
  {\bibfnamefont {Claudia}\ \bibnamefont {Felser}},\ }\bibfield  {title}
  {\enquote {\bibinfo {title} {From colossal to zero: Controlling the anomalous
  hall effect in magnetic heusler compounds via berry curvature design},}\
  }\href {\doibase 10.1103/PhysRevX.8.041045} {\bibfield  {journal} {\bibinfo
  {journal} {Phys. Rev. X}\ }\textbf {\bibinfo {volume} {8}},\ \bibinfo {pages}
  {041045} (\bibinfo {year} {2018})}\BibitemShut {NoStop}%
\bibitem [{\citenamefont {Noky}\ \emph {et~al.}(2019)\citenamefont {Noky},
  \citenamefont {Xu}, \citenamefont {Felser},\ and\ \citenamefont
  {Sun}}]{Noky2019}%
  \BibitemOpen
  \bibfield  {author} {\bibinfo {author} {\bibfnamefont {Jonathan}\
  \bibnamefont {Noky}}, \bibinfo {author} {\bibfnamefont {Qiunan}\ \bibnamefont
  {Xu}}, \bibinfo {author} {\bibfnamefont {Claudia}\ \bibnamefont {Felser}}, \
  and\ \bibinfo {author} {\bibfnamefont {Yan}\ \bibnamefont {Sun}},\ }\bibfield
   {title} {\enquote {\bibinfo {title} {Large anomalous hall and nernst effects
  from nodal line symmetry breaking in ${\mathrm{fe}}_{2}\mathrm{Mn}x$ ($x$ =
  p, as, sb)},}\ }\href {\doibase 10.1103/PhysRevB.99.165117} {\bibfield
  {journal} {\bibinfo  {journal} {Phys. Rev. B}\ }\textbf {\bibinfo {volume}
  {99}},\ \bibinfo {pages} {165117} (\bibinfo {year} {2019})}\BibitemShut
  {NoStop}%
\bibitem [{\citenamefont {Li}\ \emph {et~al.}(2020)\citenamefont {Li},
  \citenamefont {Koo}, \citenamefont {Ning}, \citenamefont {Li}, \citenamefont
  {Miao}, \citenamefont {Min}, \citenamefont {Zhu}, \citenamefont {Wang},
  \citenamefont {Alem}, \citenamefont {Liu}, \citenamefont {Mao},\ and\
  \citenamefont {Yan}}]{Li2020}%
  \BibitemOpen
  \bibfield  {author} {\bibinfo {author} {\bibfnamefont {Peigang}\ \bibnamefont
  {Li}}, \bibinfo {author} {\bibfnamefont {Jahyun}\ \bibnamefont {Koo}},
  \bibinfo {author} {\bibfnamefont {Wei}\ \bibnamefont {Ning}}, \bibinfo
  {author} {\bibfnamefont {Jinguo}\ \bibnamefont {Li}}, \bibinfo {author}
  {\bibfnamefont {Leixin}\ \bibnamefont {Miao}}, \bibinfo {author}
  {\bibfnamefont {Lujin}\ \bibnamefont {Min}}, \bibinfo {author} {\bibfnamefont
  {Yanglin}\ \bibnamefont {Zhu}}, \bibinfo {author} {\bibfnamefont
  {Yu}~\bibnamefont {Wang}}, \bibinfo {author} {\bibfnamefont {Nasim}\
  \bibnamefont {Alem}}, \bibinfo {author} {\bibfnamefont {Chao-Xing}\
  \bibnamefont {Liu}}, \bibinfo {author} {\bibfnamefont {Zhiqiang}\
  \bibnamefont {Mao}}, \ and\ \bibinfo {author} {\bibfnamefont {Binghai}\
  \bibnamefont {Yan}},\ }\bibfield  {title} {\enquote {\bibinfo {title} {Giant
  room temperature anomalous hall effect and tunable topology in a
  ferromagnetic topological semimetal co2mnal},}\ }\href {\doibase
  10.1038/s41467-020-17174-9} {\bibfield  {journal} {\bibinfo  {journal} {Nat.
  Commun.}\ }\textbf {\bibinfo {volume} {11}},\ \bibinfo {pages} {3476}
  (\bibinfo {year} {2020})}\BibitemShut {NoStop}%
\bibitem [{\citenamefont {Minami}\ \emph {et~al.}(2020)\citenamefont {Minami},
  \citenamefont {Ishii}, \citenamefont {Hirayama}, \citenamefont {Nomoto},
  \citenamefont {Koretsune},\ and\ \citenamefont {Arita}}]{Minami2020}%
  \BibitemOpen
  \bibfield  {author} {\bibinfo {author} {\bibfnamefont {Susumu}\ \bibnamefont
  {Minami}}, \bibinfo {author} {\bibfnamefont {Fumiyuki}\ \bibnamefont
  {Ishii}}, \bibinfo {author} {\bibfnamefont {Motoaki}\ \bibnamefont
  {Hirayama}}, \bibinfo {author} {\bibfnamefont {Takuya}\ \bibnamefont
  {Nomoto}}, \bibinfo {author} {\bibfnamefont {Takashi}\ \bibnamefont
  {Koretsune}}, \ and\ \bibinfo {author} {\bibfnamefont {Ryotaro}\ \bibnamefont
  {Arita}},\ }\bibfield  {title} {\enquote {\bibinfo {title} {Enhancement of
  the transverse thermoelectric conductivity originating from stationary points
  in nodal lines},}\ }\href {\doibase 10.1103/PhysRevB.102.205128} {\bibfield
  {journal} {\bibinfo  {journal} {Phys. Rev. B}\ }\textbf {\bibinfo {volume}
  {102}},\ \bibinfo {pages} {205128} (\bibinfo {year} {2020})}\BibitemShut
  {NoStop}%
\bibitem [{\citenamefont {Marzari}\ and\ \citenamefont
  {Vanderbilt}(1997)}]{Marzari1997}%
  \BibitemOpen
  \bibfield  {author} {\bibinfo {author} {\bibfnamefont {Nicola}\ \bibnamefont
  {Marzari}}\ and\ \bibinfo {author} {\bibfnamefont {David}\ \bibnamefont
  {Vanderbilt}},\ }\bibfield  {title} {\enquote {\bibinfo {title} {Maximally
  localized generalized wannier functions for composite energy bands},}\ }\href
  {\doibase 10.1103/PhysRevB.56.12847} {\bibfield  {journal} {\bibinfo
  {journal} {Phys. Rev. B}\ }\textbf {\bibinfo {volume} {56}},\ \bibinfo
  {pages} {12847--12865} (\bibinfo {year} {1997})}\BibitemShut {NoStop}%
\bibitem [{\citenamefont {Weng}\ \emph {et~al.}(2009)\citenamefont {Weng},
  \citenamefont {Ozaki},\ and\ \citenamefont {Terakura}}]{Weng2009}%
  \BibitemOpen
  \bibfield  {author} {\bibinfo {author} {\bibfnamefont {Hongming}\
  \bibnamefont {Weng}}, \bibinfo {author} {\bibfnamefont {Taisuke}\
  \bibnamefont {Ozaki}}, \ and\ \bibinfo {author} {\bibfnamefont {Kiyoyuki}\
  \bibnamefont {Terakura}},\ }\bibfield  {title} {\enquote {\bibinfo {title}
  {Revisiting magnetic coupling in transition-metal-benzene complexes with
  maximally localized wannier functions},}\ }\href {\doibase
  10.1103/PhysRevB.79.235118} {\bibfield  {journal} {\bibinfo  {journal} {Phys.
  Rev. B}\ }\textbf {\bibinfo {volume} {79}},\ \bibinfo {pages} {235118}
  (\bibinfo {year} {2009})}\BibitemShut {NoStop}%
\bibitem [{\citenamefont {Fukui}\ \emph {et~al.}(2005)\citenamefont {Fukui},
  \citenamefont {Hatsugai},\ and\ \citenamefont {Suzuki}}]{Suzuki2005}%
  \BibitemOpen
  \bibfield  {author} {\bibinfo {author} {\bibfnamefont {Takahiro}\
  \bibnamefont {Fukui}}, \bibinfo {author} {\bibfnamefont {Yasuhiro}\
  \bibnamefont {Hatsugai}}, \ and\ \bibinfo {author} {\bibfnamefont {Hiroshi}\
  \bibnamefont {Suzuki}},\ }\bibfield  {title} {\enquote {\bibinfo {title}
  {Chern numbers in discretized brillouin zone: Efficient method of computing
  (spin) hall conductances},}\ }\href {\doibase 10.1143/JPSJ.74.1674}
  {\bibfield  {journal} {\bibinfo  {journal} {J. Phys. Soc. Jpn.}\ }\textbf
  {\bibinfo {volume} {74}},\ \bibinfo {pages} {1674--1677} (\bibinfo {year}
  {2005})}\BibitemShut {NoStop}%
\bibitem [{\citenamefont {Dudarev}\ \emph {et~al.}(1998)\citenamefont
  {Dudarev}, \citenamefont {Botton}, \citenamefont {Savrasov}, \citenamefont
  {Humphreys},\ and\ \citenamefont {Sutton}}]{Dudarev1998}%
  \BibitemOpen
  \bibfield  {author} {\bibinfo {author} {\bibfnamefont {S.~L.}\ \bibnamefont
  {Dudarev}}, \bibinfo {author} {\bibfnamefont {G.~A.}\ \bibnamefont {Botton}},
  \bibinfo {author} {\bibfnamefont {S.~Y.}\ \bibnamefont {Savrasov}}, \bibinfo
  {author} {\bibfnamefont {C.~J.}\ \bibnamefont {Humphreys}}, \ and\ \bibinfo
  {author} {\bibfnamefont {A.~P.}\ \bibnamefont {Sutton}},\ }\bibfield  {title}
  {\enquote {\bibinfo {title} {Electron-energy-loss spectra and the structural
  stability of nickel oxide: An {LSDA}+{U} study},}\ }\href {\doibase
  10.1103/PhysRevB.57.1505} {\bibfield  {journal} {\bibinfo  {journal} {Phys.
  Rev. B}\ }\textbf {\bibinfo {volume} {57}},\ \bibinfo {pages} {1505--1509}
  (\bibinfo {year} {1998})}\BibitemShut {NoStop}%
\bibitem [{\citenamefont {Haule}\ \emph {et~al.}(2010)\citenamefont {Haule},
  \citenamefont {Yee},\ and\ \citenamefont {Kim}}]{Haule2010}%
  \BibitemOpen
  \bibfield  {author} {\bibinfo {author} {\bibfnamefont {Kristjan}\
  \bibnamefont {Haule}}, \bibinfo {author} {\bibfnamefont {Chuck-Hou}\
  \bibnamefont {Yee}}, \ and\ \bibinfo {author} {\bibfnamefont {Kyoo}\
  \bibnamefont {Kim}},\ }\bibfield  {title} {\enquote {\bibinfo {title}
  {Dynamical mean-field theory within the full-potential methods: Electronic
  structure of {C}e{I}r{I}n$_5$, {C}e{C}o{I}n$_5$, and {C}e{R}h{I}n$_5$},}\
  }\href {\doibase 10.1103/PhysRevB.81.195107} {\bibfield  {journal} {\bibinfo
  {journal} {Phys. Rev. B}\ }\textbf {\bibinfo {volume} {81}},\ \bibinfo
  {pages} {195107} (\bibinfo {year} {2010})}\BibitemShut {NoStop}%
\bibitem [{\citenamefont {Haule}(2018)}]{Haule2018}%
  \BibitemOpen
  \bibfield  {author} {\bibinfo {author} {\bibfnamefont {Kristjan}\
  \bibnamefont {Haule}},\ }\bibfield  {title} {\enquote {\bibinfo {title}
  {Structural predictions for correlated electron materials using the
  functional dynamical mean field theory approach},}\ }\href {\doibase
  10.7566/JPSJ.87.041005} {\bibfield  {journal} {\bibinfo  {journal} {J. Phys.
  Soc. Jpn.}\ }\textbf {\bibinfo {volume} {87}},\ \bibinfo {pages} {041005}
  (\bibinfo {year} {2018})}\BibitemShut {NoStop}%
\bibitem [{\citenamefont {Deb}\ \emph {et~al.}(2007)\citenamefont {Deb},
  \citenamefont {Itou}, \citenamefont {Tsurkan},\ and\ \citenamefont
  {Sakurai}}]{Sakurai2007}%
  \BibitemOpen
  \bibfield  {author} {\bibinfo {author} {\bibfnamefont {Aniruddha}\
  \bibnamefont {Deb}}, \bibinfo {author} {\bibfnamefont {M.}~\bibnamefont
  {Itou}}, \bibinfo {author} {\bibfnamefont {V.}~\bibnamefont {Tsurkan}}, \
  and\ \bibinfo {author} {\bibfnamefont {Y.}~\bibnamefont {Sakurai}},\
  }\bibfield  {title} {\enquote {\bibinfo {title} {Effect of substitution of
  {C}l and {B}r for {S}e in the ferromagnetic spinel {C}u{C}r$_2${S}e$_4$: A
  magnetic compton profile study},}\ }\href {\doibase
  10.1103/PhysRevB.75.024413} {\bibfield  {journal} {\bibinfo  {journal} {Phys.
  Rev. B}\ }\textbf {\bibinfo {volume} {75}},\ \bibinfo {pages} {024413}
  (\bibinfo {year} {2007})}\BibitemShut {NoStop}%
\bibitem [{\citenamefont {Ceperley}\ and\ \citenamefont
  {Alder}(1980)}]{Ceperley1980}%
  \BibitemOpen
  \bibfield  {author} {\bibinfo {author} {\bibfnamefont {D.~M.}\ \bibnamefont
  {Ceperley}}\ and\ \bibinfo {author} {\bibfnamefont {B.~J.}\ \bibnamefont
  {Alder}},\ }\bibfield  {title} {\enquote {\bibinfo {title} {Ground state of
  the electron gas by a stochastic method},}\ }\href {\doibase
  10.1103/PhysRevLett.45.566} {\bibfield  {journal} {\bibinfo  {journal} {Phys.
  Rev. Lett.}\ }\textbf {\bibinfo {volume} {45}},\ \bibinfo {pages} {566--569}
  (\bibinfo {year} {1980})}\BibitemShut {NoStop}%
\bibitem [{\citenamefont {Sugano}\ \emph {et~al.}((Academic, New York,
  1970))\citenamefont {Sugano}, \citenamefont {Tanabe},\ and\ \citenamefont
  {Kamimura}}]{Kamimura1970}%
  \BibitemOpen
  \bibfield  {author} {\bibinfo {author} {\bibfnamefont {S.}~\bibnamefont
  {Sugano}}, \bibinfo {author} {\bibfnamefont {Y.}~\bibnamefont {Tanabe}}, \
  and\ \bibinfo {author} {\bibfnamefont {H.}~\bibnamefont {Kamimura}},\
  }\href@noop {} {\emph {\bibinfo {title} {Multiplets of Transition-Metal Ions
  in Crystals}}}\ (\bibinfo {year} {(Academic, New York, 1970)})\BibitemShut
  {NoStop}%
\end{thebibliography}%

\end{document}